\documentclass[journal]{IEEEtran}
\usepackage{epsfig,graphics,epsf,color,amsmath,balance,cite,amsfonts,multirow,stfloats,algorithm,algorithmic,mathrsfs,amssymb,color,subfigure}
\usepackage[dvipdfm,unicode,colorlinks, linkcolor=black,anchorcolor=black,citecolor=black]{hyperref}
\usepackage{url,makecell}
\newtheorem{theorem}{Theorem}
\newtheorem{lemma}{Lemma}
\newtheorem{proposition}{Proposition}

\begin{document}

\title{MSE-Based Training and Transmission Optimization for MIMO ISAC Systems}

\author{Zhenyao He,~\IEEEmembership{Student Member,~IEEE,}~Wei Xu,~\IEEEmembership{Senior Member,~IEEE,}~Hong Shen,~\IEEEmembership{Member,~IEEE,}\\~Yonina C. Eldar,~\IEEEmembership{Fellow,~IEEE,} and Xiaohu You,~\IEEEmembership{Fellow,~IEEE}

\thanks{Zhenyao He, Wei Xu, Hong Shen, and Xiaohu You are with the National Mobile Communications Research Laboratory, Southeast University, Nanjing 210096, China (e-mail: \{hezhenyao, wxu, shhseu, xhyu\}@seu.edu.cn).

Yonina C. Eldar is with the Faculty of Mathematics and Computer Science, Weizmann Institute of Science, Rehovot 7610001, Israel (e-mail: yonina.eldar@weizmann.ac.il).}

}
\maketitle
\begin{abstract}
In this paper, we investigate a multiple-input multiple-output (MIMO) integrated sensing and communication (ISAC) system
under typical block-fading channels.
As a non-trivial extension to most existing works on ISAC, both the training and transmission signals sent by the ISAC transmitter are exploited for sensing. Specifically, we develop two training and transmission design schemes to minimize a weighted sum of the mean-squared errors (MSEs) of data transmission and radar target response matrix (TRM) estimation.
For the former, we first optimize the training signal for simultaneous communication channel and radar TRM estimation. Then, based on the estimated instantaneous channel state information (CSI), we propose an efficient majorization-minimization (MM)-based robust ISAC transmission design, where a semi-closed form solution is obtained in each iteration.
For the second scheme, the ISAC transmitter is assumed to have statistical CSI only for reducing the feedback overhead.
With CSI statistics available, we integrate the training and transmission design into one single problem and propose an MM-based alternating algorithm to find a high-quality solution. In addition, we provide alternative structured and low-complexity solutions for both schemes under certain special cases.
Finally, simulation results demonstrate that the radar performance is significantly improved compared to the existing scheme that integrates sensing into the transmission stage only.
Moreover, it is verified that the investigated two schemes have advantages in terms of communication and sensing performances, respectively.
\end{abstract}

\begin{IEEEkeywords}
Integrated sensing and communication (ISAC), multiple-input multiple-output (MIMO) channel estimation, robust transmission, joint training and transmission optimization, mean-squared error (MSE).
\end{IEEEkeywords}

\section{Introduction}
Integrated sensing and communication (ISAC) has become an emerging and promising technique for future wireless networks, aiming to realize the integration of reliable communications with high-accuracy sensing \cite{overviewW.Xu,Z.HeNFISAC}. Compared to conventional separate communication and radar systems, ISAC can reduce implementation costs and enhance spectral efficiency by sharing the hardware platform and spectrum \cite{overviewJ.Zhang1,overviewJ.Zhang2,overviewF.Liu}, which also brings new challenges to system design.

To achieve ISAC and fully exploit its advantages, various physical-layer transmission techniques have been proposed in recent years \cite{overviewJ.Zhang1,overviewJ.Zhang2,overviewF.Liu}. In particular, multi-antenna beamforming plays a critical role in both multiple-input multiple-output (MIMO) communication and MIMO radar systems, which has also attracted considerable research interest in MIMO ISAC.
For example, motivated by the fact that the performance of MIMO radar highly relies on the beampattern of the probing signals \cite{P.StoicaMIMOradar1}, the authors of \cite{XLiuTSP2020,X.LiuJSAC,Z.LyuArxiv,Z.HeWCL2022} performed the beamforming design in MIMO ISAC by manipulating the beampattern of the transmit signal. Concretely, the transmitted ISAC signal is designed to satisfy communication requirements and meanwhile fulfill additional restrictions on the beampattern, such as minimizing beampattern matching error \cite{XLiuTSP2020,X.LiuJSAC} and guaranteeing minimum beampattern gain \cite{Z.LyuArxiv,Z.HeWCL2022}, to ensure the sensing performance.
ISAC beamforming has also been widely studied when considering more specific sensing tasks, e.g., target detection and parameter estimation.
For example, in \cite{L.Chen2022JSAC,C.Tsinos2021JSTSP,Z.HeJSAC}, the authors investigated ISAC transceiver beamforming design for achieving simultaneous communication and target detection, where the radar receive signal-to-interference-plus-noise ratio (SINR) is chosen as the sensing performance metric, since the target detection probability is monotonically increasing with respect to the receive SINR \cite{G.CuiTSP2014}.
The authors of \cite{F.LiuTSP2021CRB} minimized the Cram\'er-Rao bound (CRB) of the target parameter estimation for MIMO ISAC, while ensuring the communication SINR.
These works all assume availability of accurate instantaneous channel state information (CSI) through a dedicated channel estimation process.

\begin{table*}[t]
\caption{Comparisons With Existing Related ISAC Works}
    \centering
	\begin{tabular}{|c|c|c|c|c|c|}
\hline  \multirow{2}*{\textbf{Study}} & \textbf{Reuse training} & \textbf{Training}  & \textbf{Transmission}  & \textbf{Joint}  & \multirow{2}*{ \textbf{Main contribution}} \\
          & \textbf{signal for sensing} & \textbf{design} & \textbf{design}  & \textbf{design} & \\
\hline  \cite{XLiuTSP2020,X.LiuJSAC,Z.LyuArxiv,Z.HeWCL2022,L.Chen2022JSAC,C.Tsinos2021JSTSP,Z.HeJSAC,G.CuiTSP2014,F.LiuTSP2021CRB}  && & $\ast$ &  &  \makecell[l]{Optimize ISAC transmission with accurate CSI availability} \\
\hline  \cite{robustISAC1,robustISAC2,robustISAC3,robustISAC4}  && & $\ast$ &  &  \makecell[l]{Investigate robust ISAC transmission with imperfect CSI} \\
\hline \cite{Spatio-temporal} & $\ast$ &  & $\ast$ & &\makecell[l]{Fix orthogonal training signal and optimize ISAC transmission}\\
\hline  \cite{MengHua} & $\ast$ &  & & $\ast$ & \makecell[l]{Jointly optimize training and transmission with a single-antenna user} \\
\hline  This work & $\ast$ & $\ast$ &$\ast$ & $\ast$ & \makecell[l]{Systematically study MIMO ISAC training and transmission design}\\
\hline
	\end{tabular}%
\end{table*}\label{table:comparison}

In fact, CSI acquisition is an important task for MIMO communications and MIMO ISAC. Owing to the presence of noise and limited pilots, perfect CSI is difficult to obtain in practice. Moreover, the CSI feedback overhead is also a detrimental factor for frequency-division duplex (FDD) MIMO systems.
Various possible solutions have been proposed to address these issues, which can be classified into the following three aspects: 1) training signal design for improving channel estimation accuracy; 2) robust transmission design for alleviating the impact of imperfect CSI; 3) statistical CSI-based design to avoid the high-overhead instantaneous CSI feedback.
The literature review for these aspects in both MIMO communications and MIMO ISAC are elaborated in the sequel.

Training signal design has been widely investigated in MIMO communications \cite{M.BigueshTSP2006Training-based,MIMOCE3}. However, to avoid the performance degradation of the communication channel estimation, in most existing ISAC literature the training signal is utilized to estimate the communication channel only, without considering the possible incorporation of sensing.
Inspired by the similar goals of radar target response matrix (TRM) and communication channel estimation, it is natural to consider integrating the sensing functionality into the training stage \cite{Spatio-temporal,MengHua}.
In \cite{Spatio-temporal}, the authors reused the orthogonal training signal for sensing and optimized the transmission signal to maximize a weighted sum of communication and sensing mutual information (MI). The authors of \cite{MengHua} optimized both the training and transmission signals to maximize the radar MI, subject to the constraints of communication SINR and power limitation.

Robust beamforming has been studied to relieve the impact of imperfect CSI on MIMO communications \cite{robust1,C.Xingmonotonic}. The extension to ISAC has been studied in \cite{robustISAC1,robustISAC2,robustISAC3,robustISAC4}. Concretely, the authors of \cite{robustISAC1} maximized the radar output power in the direction of the target under a series of probabilistic outage constraints for all communication users in the presence of CSI errors. In \cite{robustISAC2,robustISAC3,robustISAC4}, the authors investigated robust beamforming optimization from the perspective of physical layer security in ISAC, with the consideration of different kinds of CSI uncertainties for both legitimate users and eavesdroppers. However, these works only studied transmission design, without considering the possible combination of training and transmission designs.

For the purpose of reducing CSI feedback overhead, many works investigated statistical CSI-based MIMO transmission design due to the slow time-varying feature of the channel statistics \cite{partialCSI1,partialCSI2,jointdesign1}.
In this case, a joint design of training and transmission can be carried out towards a long-term average performance independent of the instantaneous CSI \cite{jointdesign2,C.Xingjointdesign}.
However, the channel statistics-based joint training and transmission design in the context of ISAC has seldom been considered except for a recent work \cite{MengHua}, where the training and transmission signals are jointly designed for radar MI maximization. Nevertheless, a relatively simple scenario with only one single-antenna user was studied in \cite{MengHua}.

In this paper we investigate the training and transmission optimization in a MIMO ISAC system, which consists of an ISAC transmitter, a colocated radar receiver, and a communication receiver.
Our main contributions are summarized as follows and a comparison with previous related ISAC works is summarized in Table~\ref{table:comparison}.
\begin{itemize}
\item As an extension to most existing works on ISAC, e.g., \cite{XLiuTSP2020,X.LiuJSAC,Z.LyuArxiv,Z.HeWCL2022,L.Chen2022JSAC,C.Tsinos2021JSTSP,Z.HeJSAC,G.CuiTSP2014,F.LiuTSP2021CRB,robustISAC1,robustISAC2,robustISAC3,robustISAC4}, we exploit both the training and transmission signals sent by the ISAC transmitter for sensing, resulting in a longer radar sequence and better sensing performance.
\item  For communication, the training and transmission signals are, respectively, utilized for channel estimation and data transmission. For sensing, the two signals are combined into a long sequence reflected and received by the radar receiver for TRM estimation. Towards the diverse demands, we propose two novel design schemes with different levels of CSI knowledge available at the ISAC transmitter for the corresponding training and transmission design, which are beneficial for communication and sensing, respectively.
  \item First, we investigate an instantaneous CSI-based design, where the communication receiver estimates the channel using the training signal and then feeds back the noisy CSI to the ISAC transmitter for subsequent transmission design. We first optimize the training signal to minimize a weighted sum of mean-squared errors (MSEs) for estimating communication channel and radar TRM. Then, using the CSI estimate, a robust MSE-based ISAC transmit beamforming optimization problem is formulated. We propose a majorization-minimization (MM)-based iterative algorithm for this intractable problem, where a semi closed-form optimal solution is obtained in each iteration.
  \item Second, to reduce the overhead caused by instantaneous CSI feedback, we formulate a channel statistics-based problem to jointly optimize the training and transmission, where the design goal is to minimize a weighted sum of long-term communication MSE and TRM estimation MSE. To obtain a high-quality solution, we develop an MM-based alternating optimization (AO) algorithm with guaranteed convergence.
\item We consider two special cases under which the optimal structures of the solutions to the above two schemes can be obtained. Then, the original complicated optimization problems reduce to power allocation problems which can be solved based on geometric program (GP). In addition, we also extend the proposed MSE-based ISAC training and transmission design methods to the MI maximization criterion.
  \item We perform extensive numerical simulations for performance evaluations. It is demonstrated that the sensing performance in our considered system can be improved due to the longer sequence used for radar. Moreover, it is validated that the instantaneous CSI-based and channel statistics-based schemes can achieve better communication and sensing performances, respectively.
%      the simulations also shown the inherent communication-sensing trade-off in ISAC and the trade-off between the overhead of CSI feedback and the achieved performances.
\end{itemize}

The rest of this paper is organized as follows. In Section~II, we present the system model of the considered MIMO ISAC system. Section~III and Section~IV investigate the proposed two training and transmission designs relying on instantaneous CSI and statistical CSI, respectively.
In Section~V, we develop structured and low-cost solutions under special cases.
In Section~VI, we extend the proposed MSE-based design methods to the MI maximization criterion.
In Section~VII, we evaluate the performance of the proposed algorithms through numerical simulations. Conclusions are drawn in Section~VIII.

\textit{Notations:}
Boldface lower-case and boldface upper-case letters represent vectors and matrices, respectively. Superscripts $(\cdot)^T$ and $(\cdot)^H$ stand for the transpose and the Hermitian transpose, respectively. Let $\text{Tr}(\cdot)$ denote the trace of a matrix, $\|\cdot\|$ and $|\cdot|$ denote the $\ell_2$ norm of a vector and the absolute value of a scalar, respectively.
We use $\mathbb E\{ \cdot \}$ to represent the expectation operation. Let $\mathcal R\{\cdot \}$ return the real part of a complex-valued number, $\mathbb C$ denote the set of complex-value numbers, and $\mathbf I_{N}$ stand for the identity matrix of size $N \times N$. We use $\text{vec}(\cdot)$ to denote the vectorization operation.
$\mathbf X \succeq \mathbf 0$ implies that $\mathbf X$ is positive semidefinite, $\mathbf X^{1/2}$ is the Hermitian
square root of $\mathbf X$, and $\lambda_\text{max}(\mathbf X)$ denotes the largest eigenvalue of $\mathbf X$.
Finally, $\mathcal O(\cdot)$ denotes the big-O computational complexity notation.

\section{System Model}
\begin{figure}[t]
\begin{center}
      \epsfxsize=7.0in\includegraphics[scale=0.4]{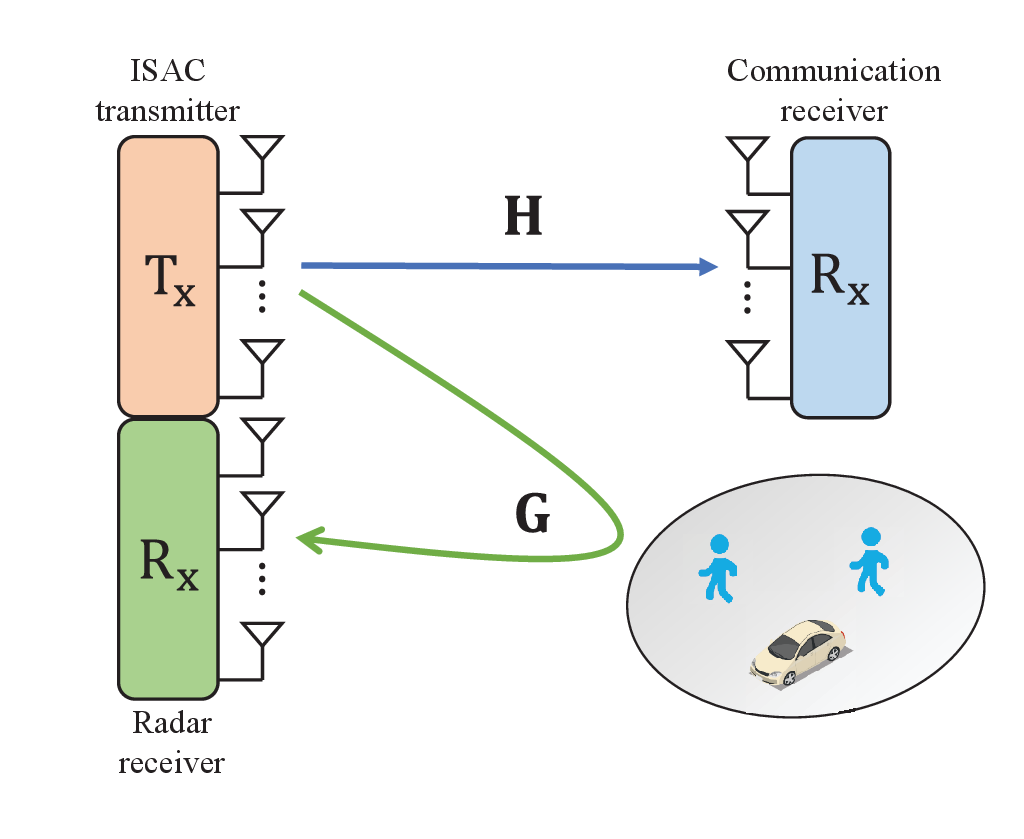}
      \caption{Diagram of the considered MIMO ISAC system.}\label{fig:model}
    \end{center}
\end{figure}

\begin{figure}[t]
\begin{center}
      \epsfxsize=7.0in\includegraphics[scale=0.4]{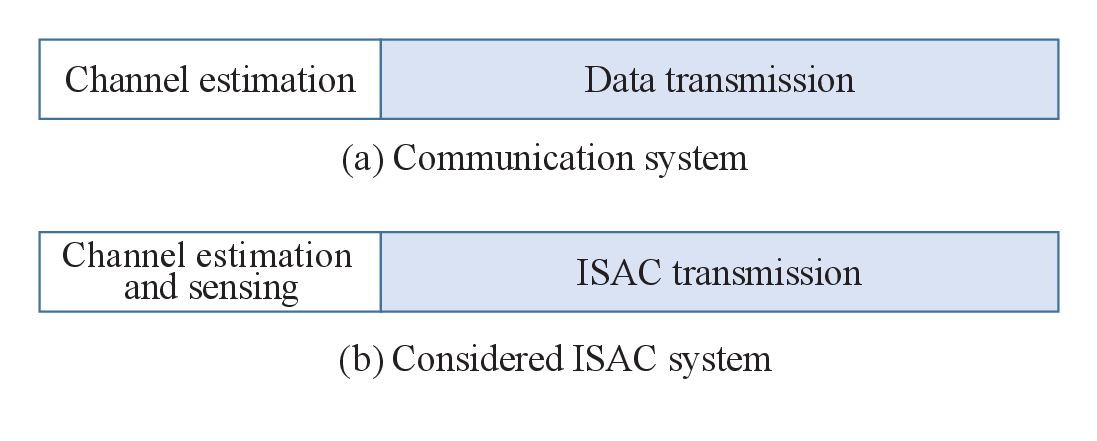}
      \caption{Frame structure over a channel fading block: (a) communication system; (b) considered ISAC system.}\label{fig:structure}
    \end{center}
\end{figure}

As shown in Fig. \ref{fig:model}, we consider a MIMO ISAC system, where an ISAC transmitter sends dual-functional signals to a communication receiver and meanwhile receives the echoes for sensing at the radar receiver. The number of antennas at the ISAC transmitter, the communication receiver, and the radar receiver are denoted by $M$, $N^\text{com}$, and $N^\text{rad}$, respectively.
Define the point-to-point MIMO communication channel by $\mathbf H \in \mathbb C^{N^\text{com} \times M}$ and the radar TRM by $\mathbf G \in \mathbb C^{N^\text{rad} \times M}$.

As depicted in Fig.~\ref{fig:structure}(a), we consider a block-fading channel model for MIMO communication, where each block is divided into training-based channel estimation and data transmission stages, respectively. In most existing ISAC works, in order to guarantee accurate CSI acquisition, the sensing functionality is combined into the data transmission stage only.
Inspired by the similar goal of sensing and communication channel estimation, i.e., recovering useful information from the wireless links, in this paper we also use the training signal for sensing, as shown in Fig.~\ref{fig:structure}(b). In other words, given the echoes of both training and data signals received at the radar receiver, the sensing task is performed with a longer radar code to achieve performance enhancement \cite{Radarbook}.
Assume that a fading block occupies $L$ time slots, which include the training symbols of length $L^\text{CE}$ and the data symbols of length $L^\text{DT}$ with $L = L^\text{CE} + L^\text{DT}$. In the sequel, we demonstrate the signal models for communication and sensing, respectively.

\subsection{Communication Model}

\subsubsection{Channel Estimation}
The ISAC transmitter sends training signals to the communication receiver for estimating the channel $\mathbf H$, which is assumed to follow the Rayleigh fading model. Specifically, denoting the training signal at slot $l$ by $\mathbf x[l] \in \mathbb C^{M \times 1}$, we express the received signal at the communication receiver, denoted by $\mathbf y[l] \in \mathbb C^{N^\text{com} \times 1}$, as
\begin{align}\label{signal:yn}
\mathbf y[l] = \mathbf H \mathbf x[l] + \mathbf z[l],\ l=1,\cdots, L^\text{CE},
\end{align}
where $\mathbf z[l] \in \mathbb C^{N^\text{com} \times 1}$ represents additive white Gaussian noise (AWGN) with covariance $\sigma^2 \mathbf I_{N^\text{com}}$.
In order to estimate $\mathbf H$ accurately, $L^\text{CE} \geq M$ training symbols are usually required \cite{M.BigueshTSP2006Training-based}. Combine the $L^\text{CE}$ received training signals and rewrite (\ref{signal:yn}) as the following compact form:
\begin{align}\label{signal:Ycom}
\mathbf Y^\text{com} = \mathbf H \mathbf X + \mathbf Z^\text{com},
\end{align}
where $\mathbf Y^\text{com} \triangleq [\mathbf y[1], \cdots, \mathbf y[L^\text{CE}] ] \in \mathbb C^{N^\text{com} \times L^\text{CE}}$, $\mathbf X \triangleq [\mathbf x[1], \cdots, \mathbf x[L^\text{CE}] ] \in \mathbb C^{M \times L^\text{CE}}$, and $\mathbf Z^\text{com} \triangleq [\mathbf z[1], \cdots, \mathbf z[L^\text{CE}] ] \in \mathbb C^{N^\text{com} \times L^\text{CE}}$ satisfying $\mathbb E\{ \text{vec}(\mathbf Z^\text{com}) \text{vec}^H(\mathbf Z^\text{com})\} = \sigma^2 \mathbf I_{L^\text{CE} N^\text{com}}$, respectively.
The training signal $\mathbf X$ satisfies a total power constraint $\text{Tr} \{\mathbf X \mathbf X^H\} \leq P^\text{CE}$ with $P^\text{CE}$ being the maximum transmit power for training.

By employing linear minimum MSE (LMMSE) channel estimation on (\ref{signal:Ycom}), the estimated channel $\mathbf {\hat H}$ equals \cite{M.BigueshTSP2006Training-based}
\begin{align}\label{def:H_hat}
\mathbf {\hat H} = \mathbf Y^\text{com} \left(\mathbf X^H \mathbf R_{\mathbf H}  \mathbf X + \sigma^2 \mathbf I_{L^\text{CE}}\right)^{-1}\mathbf X^H \mathbf R_{\mathbf H},
\end{align}
where $\mathbf R_{\mathbf H} \triangleq \frac{1}{N^\text{com}}\mathbb{E}\left\{ \mathbf H^H \mathbf H \right\} \in \mathbb C^{M \times M}$ denotes the transmit correlation,
which is previously estimated and known to the system. The channel estimate $\mathbf {\hat H}$ and the true channel $\mathbf H$ satisfy
\begin{align}
\mathbf H = \mathbf {\hat H} + \mathbf \Delta,
\end{align}
where $\mathbf \Delta$ denotes the estimation error with zero mean. The covariance matrices of $\mathbf \Delta$ and $\mathbf {\hat H}$ are, respectively, given by
\begin{align}
\mathbf R_{\mathbf \Delta} \triangleq&\ \frac{1}{N^\text{com}} \mathbb{E}\{ \mathbf \Delta^H \mathbf \Delta \} = \left(\mathbf R_{\mathbf H}^{-1} + \frac{1}{\sigma^2} \mathbf X\mathbf X^H\right)^{-1}, \nonumber \\
\mathbf R_{\mathbf {\hat H}} \triangleq&\ \frac{1}{N^\text{com}} \mathbb{E}\{ \mathbf {\hat H}^H \mathbf {\hat H} \} = \mathbf R_{\mathbf H} - \mathbf R_{\mathbf \Delta}.
\end{align}
The channel estimation MSE is
\begin{align}\label{MSE:CE}
\text{MSE}^\text{CE} \triangleq \text{Tr}\{\mathbf R_{\mathbf \Delta}\} = \text{Tr}\left\{  \left(\mathbf R_{\mathbf H}^{-1} + \frac{1}{\sigma^2} \mathbf X\mathbf X^H\right)^{-1}  \right\}.
\end{align}

\subsubsection{Data Transmission With Instantaneous CSI Feedback}
Once channel estimation is accomplished, the communication receiver feeds back $\mathbf {\hat H}$ to the transmitter via an error-free and low-delay feedback link. Subsequently, based on $\mathbf {\hat H}$, the ISAC transmitter takes the impact of imperfect CSI into consideration and conducts a corresponding robust design for data transmission. Specifically,
the received signal at the communication receiver is given by
\begin{align}
\mathbf y[l] =&\ \mathbf H \mathbf W \mathbf s[l] + \mathbf z[l], \nonumber \\
=&\ \mathbf {\hat H} \mathbf W \mathbf s[l] + \mathbf \Delta \mathbf W \mathbf s[l] + \mathbf z[l],\ l=1,\cdots, L^\text{DT},
\end{align}
where $\mathbf s[l]  \in \mathbb C^{D \times 1}$ is the data symbol vector in the $l$-th slot satisfying $ \mathbb{E}\{\mathbf s[l] \mathbf s^H[l]\} = \mathbf I_{D}$ with $D \leq \min\{M,N^\text{com}\}$ being the number of data streams, and $\mathbf W \in \mathbb C^{M \times D}$ denotes the transmit beamforming matrix satisfying a power constraint $\text{Tr} \{\mathbf W \mathbf W^H \} \leq P^\text{DT}$ with $P^\text{DT}$ being the power budget. At the communication receiver, a linear equalizer $\mathbf V \in \mathbb C^{N^\text{com} \times D}$ is used to recover the data symbol $\mathbf s[l]$ from $\mathbf y[l]$, which yields an estimate of $\mathbf s[l]$ by $\mathbf {\hat s}[l] = \mathbf V^H \mathbf y[l]$.
To accomplish the robust design of $\mathbf W$ and $\mathbf V$, in this paper we employ the MSE between $\mathbf s[l]$ and $\mathbf {\hat s}[l]$ as the performance metric, which is given by
\begin{align}\label{MSE_com_ori}
&\ \text{MSE}^\text{com} \nonumber \\
=&\
 \mathbb{E}_{\mathbf \Delta,\mathbf s[l],\mathbf z[l]} \left\{\| \mathbf V^H \mathbf y[l] - \mathbf s[l]\|^2  \right\} \nonumber\\
=&\  \text{Tr}\{
\mathbf V^H( \mathbf{\hat H} \mathbf W \mathbf W^H \mathbf{\hat H}^H + \sigma^2\mathbf I_{N^\text{com}})\mathbf V  \!-\! 2 \text{Re}\{\mathbf V^H\mathbf{\hat H} \mathbf W \} \!+\! \mathbf I_{D}
\} \nonumber\\
&\ + \mathbb{E}_{\mathbf \Delta} \{ \text{Tr}\{ \mathbf V^H \mathbf\Delta \mathbf W \mathbf W^H \mathbf\Delta^H \mathbf V\} \} \nonumber \\
= &\ \text{Tr}\{
\mathbf V^H( \mathbf{\hat H} \mathbf W \mathbf W^H \mathbf{\hat H}^H + (\text{Tr}\{ \mathbf W \mathbf W^H \mathbf R_{\mathbf \Delta} \} + \sigma^2)\mathbf I_{N^\text{com}})\mathbf V   \nonumber \\
&\ \quad\  - 2 \text{Re}\{\mathbf V^H\mathbf{\hat H} \mathbf W \} + \mathbf I_D
\}.
\end{align}
The third equality holds when $\frac{1}{M}\mathbb{E}\{ \mathbf H \mathbf H^H \} = \mathbf I_{N^\text{com}}$, i.e., the communication receiver is located in a highly scattering environment \cite{jointdesign1,jointdesign2,C.Xingjointdesign}. When $\frac{1}{M}\mathbb{E}\{ \mathbf H \mathbf H^H  \} \neq \mathbf I_{N^\text{com}}$, (\ref{MSE_com_ori}) is actually an upper bound on $\text{MSE}^\text{com}$ due to the inequality $ \text{Tr} \{\mathbf V^H \mathbf\Delta \mathbf W \mathbf W^H \mathbf\Delta^H \mathbf V\} \leq \text{Tr} \{ \mathbf\Delta \mathbf W \mathbf W^H \mathbf\Delta^H \}\text{Tr} \{\mathbf V^H \mathbf V\}$.
Based on (\ref{MSE_com_ori}), the optimal MMSE communication receiver takes the form $
\mathbf V^* =  ( \mathbf{\hat H} \mathbf W \mathbf W^H \mathbf{\hat H}^H + (\text{Tr}\{ \mathbf W \mathbf W^H \mathbf R_{\mathbf \Delta}\} + \sigma^2)\mathbf I_{N^\text{com}} )^{-1} \mathbf{\hat H} \mathbf W$ and the corresponding MSE is
\begin{align}\label{MSE:com}
\text{MSE}^\text{com}
= \text{Tr}\left\{ \left( \mathbf I_{D} +  \frac{\mathbf W^H \mathbf{\hat H}^H \mathbf{\hat H} \mathbf W} {\text{Tr}\{ \mathbf W \mathbf W^H \mathbf R_{\mathbf \Delta} \} + \sigma^2}  \right)^{-1}\right\}.
\end{align}
We will use (\ref{MSE:com}) to achieve robust design of the transmit beamforming matrix $\mathbf W$ in Section~III.

\subsubsection{Data Transmission With Statistical CSI Feedback}
Note that the instantaneous CSI feedback from the communication receiver to the ISAC transmitter yields high overhead costs, especially when the channel dimension becomes large. Therefore, we consider another statistical channel information-based design \cite{jointdesign2,C.Xingjointdesign}, where the communication receiver has the channel estimate $\mathbf{\hat H}$ while only the channel statistic $\mathbf R_\mathbf{H}$ is conveyed to the ISAC transmitter, which avoids the feedback of instantaneous CSI. In this case, we need to adopt the average performance metric instead of $\text{MSE}^\text{com}$ in (\ref{MSE:com}) relying on instantaneous CSI. Specifically, we take the expectation of $\text{MSE}^\text{com}$ with respect to $\mathbf{\hat H}$, which yields \cite{C.Xingjointdesign}
\begin{align}\label{def:MSEbar}
&\ \mathbb E_{\mathbf{\hat H}} \{\text{MSE}^\text{com}\}\nonumber\\
=&\ \mathbb E_{\mathbf{\hat H}} \left\{ \text{Tr}\left\{ \left( \mathbf I_{D} +  \frac{\mathbf W^H \mathbf{\hat H}^H \mathbf{\hat H} \mathbf W} {\text{Tr}\{ \mathbf W \mathbf W^H \mathbf R_{\mathbf \Delta} \} + \sigma^2} \right)^{-1}\right\} \right\} \nonumber\\
\geq&\   \text{Tr}\left\{ \left( \mathbf I_{D} +  \frac{\mathbf W^H \mathbb E_{\mathbf{\hat H}}\{\mathbf{\hat H}^H \mathbf{\hat H}\} \mathbf W} {\text{Tr}\{ \mathbf W \mathbf W^H \mathbf R_{\mathbf \Delta} \} + \sigma^2} \right)^{-1}\right\}  \nonumber\\
=&\ \text{Tr}\left\{ \left( \mathbf I_{D} +  \frac{N^\text{com} \mathbf W^H \mathbf R_\mathbf{\hat H} \mathbf W} {\text{Tr}\{ \mathbf W \mathbf W^H \mathbf R_{\mathbf \Delta} \} + \sigma^2} \right)^{-1}\right\}
\triangleq \overline{\text{MSE}}^\text{com} ,
\end{align}
where the inequality holds owing to the Jensen's inequality.
Note that the $\overline{\text{MSE}}^\text{com}$ in (\ref{def:MSEbar}) is a long-term average metric independent of the instantaneous CSI and can be regarded as an upper bound to the communication performance. Furthermore, this metric also facilitates the joint training and transmission optimization in Section~IV.

\subsection{Sensing Model}
Given the consecutively transmitted training signal $\mathbf X$ and the data signal $\mathbf W \mathbf S$, where $\mathbf S \triangleq [\mathbf s[1], \cdots, \mathbf s[L^\text{DT}]]\in\mathbb C^{D\times L^\text{DT}}$, the complete signal during the entire channel fading block is denoted by $\mathbf P = [\mathbf X, \mathbf W \mathbf S] \in\mathbb C^{M\times L}$, which is used for sensing. The echo received at the radar receiver, denoted by $\mathbf Y^\text{rad} \in \mathbb C^{N^\text{rad} \times L}$, is described as
\begin{align}\label{signal:Yrad}
\mathbf Y^\text{rad} = \mathbf G \mathbf P + \mathbf Z^\text{rad},
\end{align}
where $\mathbf Z^\text{rad} \in \mathbb C^{N^\text{rad} \times L}$ stands for the AWGN satisfying $\mathbb E\{ \text{vec}(\mathbf Z^\text{rad}) \text{vec}^H(\mathbf Z^\text{rad})\} = \sigma^2 \mathbf I_{L  N^\text{rad}}$.
In this paper, the complete response matrix $\mathbf G$ is estimated, with the knowledge of the second-order statistical information $\mathbf R_{\mathbf G} \triangleq \frac{1}{N^\text{rad}} \mathbb{E}\{ \mathbf G^H \mathbf G \} \in \mathbb C^{M \times M}$ \cite{Y.Yang2007,B.TangTSP2019,T.Naghibi2011}.
If necessary, more detailed information of the target, e.g., angle and range, can be further captured based on the TRM estimate $\mathbf {\hat G}$ \cite{F.LiuTSP2021CRB}, and a smaller TRM estimation error can improve the accuracy of target parameter estimation.
The estimation MSE of $\mathbf G$ using the LMMSE estimator is \cite{Y.Yang2007,T.Naghibi2011}
\begin{align}
\!\!\text{MSE}^\text{rad}
=&\ \frac{1}{N^\text{rad}} \mathbb{E}\left\{ \| \mathbf G - \mathbf {\hat G}\|_F^2 \right\} \nonumber \\
=&\  \text{Tr}\left\{\! \left(\! \mathbf R_{\mathbf G}^{-1} \!+\! \frac{1}{\sigma^2}\! \left(\mathbf X\mathbf X^H \!+\! \mathbf W \mathbf S\mathbf S^H \mathbf W^H \right)\! \right)^{-1} \right\} \label{MSE:rad_real} \\
\approx&\ \text{Tr}\left\{\! \left(\! \mathbf R_{\mathbf G}^{-1} \!+\! \frac{1}{\sigma^2}\! \mathbf X\mathbf X^H \!+\! \frac{L^\text{DT}}{\sigma^2}\mathbf W \mathbf W^H \right)^{-1} \right\}, \label{MSE:rad}
\end{align}
where the last step follows from the independence among zero-mean random data, i.e., $ \mathbf S \mathbf S^H \approx L^\text{DT} \mathbf I_D$.
We note that the above approximation has also been adopted in prior ISAC works \cite{X.LiuJSAC,F.LiuTSP2021CRB} and the accuracy of this approximation will be validated via simulation in Section~\ref{section:simulation}.

In this paper, $\text{MSE}^\text{com}$ of data communication and $\text{MSE}^\text{rad}$ of radar TRM estimation are the performance metrics for the considered ISAC design.
We aim at jointly optimizing the training signal $\mathbf X$ and the transmit beamforming $\mathbf W$ to minimize a weighted sum of $\text{MSE}^\text{com}$ and $\text{MSE}^\text{rad}$.
It is worth noting that, from the communication perspective, $\mathbf X$ and $\mathbf W$ generally need to be separately optimized to minimize the channel estimation MSE \cite{M.BigueshTSP2006Training-based,MIMOCE3} and the data transmission MSE \cite{robust1,C.Xingmonotonic}, respectively. While for the radar system, $\mathbf X$ and $\mathbf W$ need to be jointly designed to minimize the sensing MSE \cite{Y.Yang2007,T.Naghibi2011}.
Therefore, for the considered ISAC system, we develop two schemes to optimize $\mathbf X$ and $\mathbf W$ in both sequential and joint manners, which correspond to the communication-preferred and the radar-preferred ISAC designs and are presented in Sections~III and IV, respectively.

\section{Sequential Design With Instantaneous CSI Feedback}
In this section, we optimize the training and transmission signals in a sequential manner with instantaneous CSI feedback.
Concretely, the ISAC training signal $\mathbf X$ is designed and sent first. Then, the communication receiver performs channel estimation based on the optimized $\mathbf X$ and feeds back the channel estimate $\mathbf{\hat H}$ to the ISAC transmitter. Subsequently, a robust ISAC transmit beamforming design is conducted based on $\mathbf{\hat H}$. This scheme is expected to achieve better communication performance at a cost of CSI feedback overhead.

\subsection{Problem Formulation}
During the training stage, the signal $\mathbf X$ is used to simultaneously recover the communication channel $\mathbf H$ and the radar TRM $\mathbf G$. Hence, we formulate the ISAC training design problem by minimizing a weighted sum of the corresponding estimation MSEs, which takes the form:
\begin{align}\label{prob:X}
(\mathcal P_1):\
 \mathop \text{minimize}\limits_{\mathbf X} \quad &
\frac{\omega_1}{M}  \text{Tr}\left\{\left(\mathbf R_{\mathbf H}^{-1} + \mathbf X \mathbf X^H/\sigma^2 \right)^{-1}\right\} \nonumber \\
& + \frac{\omega_2 }{M} \text{Tr}\left\{\left(\mathbf R_{\mathbf G}^{-1} + \mathbf X \mathbf X^H/\sigma^2 \right)^{-1}\right\} \nonumber \\
\text{subject to} \quad & \text{Tr}\left\{ \mathbf X \mathbf X^H\right\} \leq P^\text{CE},
\end{align}
where the MSEs are normalized by the number of transmit antennas $M$ and $\omega_1 \in [0,1]$ and $\omega_2 = 1 - \omega_1$ are the weighting factors determining the trade-off between communication and sensing.

Subsequently, with the instantaneous $\mathbf{\hat H}$ fed back from the communication receiver, an ISAC beamforming design problem is formulated for simultaneous robust data transmission and target sensing during the transmission stage, which is given by
\begin{align}\label{prob:W}
\!\!\!\!(\mathcal P_2):\
\mathop \text{minimize}\limits_{\mathbf W} \quad &
\frac{\omega_1}{D} \text{Tr}\left\{ \left( \mathbf I_{D} +  \frac{\mathbf W^H \mathbf{\hat H}^H \mathbf{\hat H} \mathbf W} {\text{Tr}\{ \mathbf W \mathbf W^H \mathbf R_{\mathbf \Delta} \} + \sigma^2} \right)^{-1}\right\} \nonumber\\
&\!\!\!\!\!\!\!\!\!\!\!\!\!\!\!\!\!\!\!\!\!\!\!\!\!\!\!\!\!\!\!\! +  \frac{\omega_2 }{M} \text{Tr}\left\{ \left(\mathbf R_{\mathbf G}^{-1} + \mathbf X^\star(\mathbf X^\star)^H/\sigma^2 + L^\text{DT} \mathbf W \mathbf W^H/\sigma^2 \right)^{-1} \right\} \nonumber \\
\text{subject to} \quad & \text{Tr}\left\{ \mathbf W \mathbf W^H\right\} \leq P^\text{DT},
\end{align}
where $\text{MSE}^\text{com}$ is normalized by the number of data streams $D$, $\mathbf X^\star$ represents the optimized training signal achieved by solving $(\mathcal P_1)$, and $\mathbf R_{\mathbf \Delta}$ is calculated as $\mathbf R_{\mathbf \Delta} = (\mathbf R_{\mathbf H}^{-1} + \mathbf X^\star (\mathbf X^\star)^H/\sigma^2)^{-1}$.
Note that these two problems are solved once in a sequential manner, instead of optimizing one variable with the other one fixed as in an alternating manner.

These weighted sum MSE minimization problems in ISAC, especially $(\mathcal P_2)$, are intractable. In particular, $(\mathcal P_2)$ cannot be directly handled by the existing methods proposed for MIMO communications, e.g., \cite{robust1,C.Xingmonotonic}, and MIMO radar, e.g., \cite{Y.Yang2007,T.Naghibi2011}.
Towards these issues, our proposed solutions are given as follows.

\subsection{Solution for ISAC Training}
The training signal design problem in $(\mathcal P_1)$ can be readily transformed into a semidefinite program (SDP) formulation with respect to $\mathbf R_{\mathbf X} \triangleq \mathbf X \mathbf X^H$:
\begin{align}\label{prob:RX}
\mathop \text{minimize}\limits_{\mathbf R_{\mathbf X}\succeq \mathbf 0} \quad &
\frac{\omega_1}{M}  \text{Tr}\left\{\left(\mathbf R_{\mathbf H}^{-1} + \mathbf R_{\mathbf X}/\sigma^2 \right)^{-1}\right\} \nonumber \\
& + \frac{\omega_2 }{M} \text{Tr}\left\{\left(\mathbf R_{\mathbf G}^{-1} + \mathbf R_{\mathbf X}/\sigma^2 \right)^{-1}\right\} \nonumber \\
\text{subject to} \quad & \text{Tr}\left\{ \mathbf R_{\mathbf X}\right\} \leq P^\text{CE}.
\end{align}
Note that $\mathbf{X}$ denotes the deterministic training sequence with arbitrary rank. Hence, there are no rank limitations on $\mathbf{R}_{\mathbf{X}}$ either.
Problem (\ref{prob:RX}) can be solved via off-the-shelf convex optimization tools, e.g., CVX \cite{CVXtool}.
Then, the optimal $\mathbf X$, i.e., the optimal solution of $(\mathcal P_1)$, can be recovered as follows:
\begin{align}\label{decomp:R_X}
\mathbf X = \left[ \mathbf R_{\mathbf X}^{1/2}, \mathbf 0_{M \times (L^\text{CE} - M)} \right] \mathbf {\bar U},
\end{align}
where $\mathbf {\bar U}$ is an arbitrary unitary matrix of size $L^\text{CE} \times L^\text{CE}$.

\subsection{Solution for ISAC Robust Transmission}
Compared to $(\mathcal P_1)$, the transmission design $(\mathcal P_2)$ is more challenging due to the signal-dependent term $\text{Tr}\{ \mathbf W \mathbf W^H \mathbf R_{\mathbf \Delta} \}$ caused by the imperfect CSI in $\text{MSE}^\text{com}$, as well as the simultaneous involvement of the sensing MSE in the objective function.
To overcome the difficulties, we employ the popular MM algorithm \cite{Y.SunTSP2017MM}. The basic idea of the MM algorithm is to construct a sequence of locally approximate surrogate problems with simpler forms, and solve them until convergence.
By manipulating the communication and sensing MSEs, we construct a surrogate function that serves as an upper bound of the objective function of $(\mathcal P_2)$. The resulting surrogate problem is given in the following proposition.

\begin{proposition}\label{prop:MM_P2}
For $(\mathcal P_2)$, a surrogate problem used in each iteration of the MM algorithm is given by
\begin{align}\label{prob:p2MM}
\mathop \text{minimize}\limits_{\mathbf W} \quad\!\! & \text{Tr}\left\{ \! \mathbf W^H  \!\left(\frac{\omega_1}{D}\mathbf \Psi  \!+ \! \lambda\frac{\omega_2}{M} \mathbf I_M \right) \! \mathbf W \! \right\} \! - \!2 \mathcal R\{ \text{Tr}\{ \mathbf \Pi^H \mathbf W \}\}\nonumber \\
\text{subject to} \quad\!\! & \text{Tr}\{ \mathbf W \mathbf W^H\} \leq P^\text{DT},
\end{align}
where the calculations of $\mathbf \Psi \succeq \mathbf 0$, $\lambda > 0$, and $\mathbf \Pi$  are based on the solution of $\mathbf W$ obtained in the previous iteration and their specific definitions are shown in (\ref{def:Psi}), (\ref{def:lambda}), and (\ref{def:Pi}), respectively, in Appendix~\ref{proof:MM_P2}.
\end{proposition}
\begin{IEEEproof}
See Appendix~\ref{proof:MM_P2}.
\end{IEEEproof}

The advantage of utilizing the MM algorithm is that, compared to $(\mathcal P_2)$, we only need to solve the compact convex problem in (\ref{prob:p2MM}) iteratively. In particular, by analyzing the Karush-Kuhn-Tucker (KKT) conditions of (\ref{prob:p2MM}), we obtain the following optimal solution:
\begin{align}\label{def:W*}
\mathbf W^\star =
\begin{cases}
\mathbf {\widetilde W}(0) \quad & \text{if}\ \text{Tr}\{\mathbf {\widetilde W}(0) \mathbf {\widetilde W}^H(0)\} \leq P^\text{DT}\\
\mathbf {\widetilde W}(\mu^\star) \quad & \text{otherwise},
\end{cases}
\end{align}
where $\mathbf {\widetilde W}(\mu) \triangleq \left(\frac{\omega_1}{D} \mathbf \Psi + \lambda \frac{\omega_2}{M} \mathbf I_M + \mu \mathbf I_M \right)^{-1} \mathbf \Pi$ and $\mu^\star \geq 0$ represents the optimal Lagrange multiplier associated with the power constraint satisfying $\mu^\star(\text{Tr}\{ \mathbf W^\star (\mathbf W^\star)^H\} - P^\text{DT}) = 0$. When $\text{Tr}\{\mathbf {\widetilde W}(0) \mathbf {\widetilde W}^H(0)\} > P^\text{DT}$, the optimal $\mu^\star > 0$ can be found via a bisection search since $\text{Tr}\{ \mathbf {\widetilde W}(\mu) \mathbf {\widetilde W}^H(\mu) \}$ is monotonically decreasing with respect to $\mu$.

\begin{algorithm}[t]
\caption{Proposed MM-Based Algorithm for Solving $(\mathcal P_2)$}
\label{alg:MM}
\begin{algorithmic}[1]
\STATE \textit{Initialization:} Set $i=0$ and initialize a feasible $ \mathbf W^{(0)}$.
\REPEAT
    \STATE Set $i=i + 1$.\\
    \STATE
    Based on $\mathbf W^{(i-1)}$, update $\mathbf W^{(i)}$ according to (\ref{def:W*}). \\
\UNTIL \textit{convergence}.
\end{algorithmic}
\end{algorithm}

The proposed MM-based algorithm for solving $(\mathcal P_2)$ is summarized in Algorithm~\ref{alg:MM}, whose convergence property is given as follows.
\begin{theorem}\label{theorem:MMconvergence}
Algorithm~\ref{alg:MM} can converge to a stationary point of $(\mathcal P_2)$.
\end{theorem}
\begin{IEEEproof}
See Appendix~\ref{proof:MMconvergence}.
\end{IEEEproof}
Furthermore, an acceleration scheme can be employed to enhance the convergence speed \cite{MMAcce1,Z.HeWCL}. The computational complexity of Algorithm~\ref{alg:MM} mainly lies in calculating (\ref{def:W*}), which has an order of $\mathcal O(M^3 + (N^\text{com})^3)$.

Note that the sequential design proposed in this section can also be applied to a time-division duplex (TDD) system, where the uplink ISAC training signal serves the dual-functional purposes of both communication channel estimation and radar TRM estimation.

\section{Joint Training and Transmission Design}
In this section, we study the joint training and transmission design which requires only statistical CSI at the ISAC transmitter, avoiding instantaneous feedback. The joint design is formulated as one single problem, which, however, involves tight variable coupling and is difficult to solve. Different from the separate design in Section~III, this scheme is expected to achieve better sensing performance.

\subsection{Problem Formulation}
With the average communication MSE in (\ref{def:MSEbar}), we are able to conduct one single joint optimization problem with respect to the training and transmission signals based on the statistical CSI, instead of the simultaneous specific estimation for each channel realization.
Accordingly, the problem is formulated as
\begin{align}\label{prob:JD}
\!\!\!(\mathcal P_3):\
\mathop \text{minimize}\limits_{\mathbf X,\mathbf W} \quad\!\! &
\frac{\omega_1}{D} \text{Tr}\left\{ \left( \mathbf I_{D} +  \frac{N^\text{com} \mathbf W^H \mathbf R_\mathbf{\hat H} \mathbf W} {\text{Tr}\{ \mathbf W \mathbf W^H \mathbf R_{\mathbf \Delta} \} + \sigma^2} \right)^{-1}\right\} \nonumber\\
&\!\!+\!\! \frac{\omega_2 }{M} \text{Tr}\left\{ \left(\mathbf R_{\mathbf G}^{-1} \!\!+\!\! \frac{1}{\sigma^2} \mathbf X\mathbf X^H \!\!+\!\! \frac{L^\text{DT}}{\sigma^2} \mathbf W \mathbf W^H \right)^{-1} \!\! \right\} \nonumber \\
\text{subject to} \quad\!\! & \mathbf R_{\mathbf \Delta} = \left(\mathbf R_{\mathbf H}^{-1} + \frac{1}{\sigma^2} \mathbf X\mathbf X^H\right)^{-1}, \nonumber \\
& \mathbf R_{\mathbf {\hat H}} = \mathbf R_{\mathbf H} - \mathbf R_{\mathbf \Delta}, \nonumber \\
& \text{Tr}\left\{ \mathbf X \mathbf X^H\right\} \leq P^\text{CE}, \nonumber \\
& \text{Tr}\left\{ \mathbf W \mathbf W^H\right\} \leq P^\text{DT},
\end{align}
where $\mathbf R_{\mathbf \Delta}$ and $\mathbf R_{\mathbf {\hat H}}$ are not optimization variables but included here for presentation brevity.
Compared to the separate design problems $(\mathcal P_1)$ and $(\mathcal P_2)$ in the previous section, $(\mathcal P_3)$ is more challenging to solve, since it not only has a nonconvex form but also includes tightly coupled variables.
Moreover, due to the simultaneous involvement of sensing MSE, the existing joint design methods proposed for MIMO communications \cite{jointdesign2,C.Xingjointdesign} are not applicable to this problem.
To address this issue, we utilize the AO framework to iteratively update one variable with the other one fixed.

\subsection{Optimization With Respect to the Training Signal}
We first concentrate on optimization with respect to the training signal $\mathbf X$.
Note that even for fixed $\mathbf W$, the subproblem of $\mathbf X$ remains intractable.
By introducing real-valued $t \geq 0, u \geq 0$, and a positive semidefinite matrix $\mathbf \Gamma \in \mathbb C^{M\times M}$, we first obtain an equivalent form of the subproblem of $\mathbf X$.
\begin{proposition}\label{prop:prob_equ}
The subproblem of $\mathbf X$ can be recast as:
\begin{align}\label{prob:JD_Xtu}
\mathop \text{minimize}\limits_{\mathbf R_{\mathbf X},\mathbf \Gamma, t,u } \ \ &
\frac{\omega_1}{D} tu  \!+\! \frac{\omega_2 }{M} \text{Tr}\left\{ \left(\mathbf R_{\mathbf G}^{-1} \!\!+\!\! \frac{1}{\sigma^2} \mathbf R_{\mathbf X} \!\!+\!\! \frac{L^\text{DT}}{\sigma^2} \mathbf W \mathbf W^H \right)^{-1} \right\} \nonumber \\
\text{subject to} \ \ & t\geq \text{Tr}\{ \mathbf W \mathbf W^H  \mathbf \Gamma \} + \sigma^2, \nonumber \\
& u \geq \text{Tr}\left\{ \left(t\mathbf I_{D}  +  N^\text{com} \mathbf W^H (\mathbf R_{\mathbf H} - \mathbf \Gamma) \mathbf W \right)^{-1}\right\}, \nonumber \\
& \mathbf \Gamma \succeq  (\mathbf R_{\mathbf H}^{-1} + \mathbf R_{\mathbf X} /\sigma^2)^{-1}  , \nonumber \\
& \text{Tr}\left\{ \mathbf R_{\mathbf X}\right\} \leq P^\text{CE}, \nonumber \\
& t \geq 0,\ u \geq 0,\ \mathbf R_{\mathbf X} \succeq \mathbf 0,\ \mathbf \Gamma \succeq \mathbf 0,
%\ \mathbf R_{\mathbf H} \succeq \mathbf \Gamma,
\end{align}
where $\mathbf R_{\mathbf X} = \mathbf X \mathbf X^H$.
\end{proposition}
\begin{IEEEproof}
See Appendix~\ref{proof:prob_equ}.
\end{IEEEproof}

For the reformulation in (\ref{prob:JD_Xtu}), it is seen that all the constraints are in convex form.
%Particularly, the third matrix inequality regarding $\mathbf \Gamma$ can be converted into the following linear matrix inequality form by utilizing the Schur's complements \cite{cvxbook}:
%\begin{align}\label{LMI:Gamma}
%\begin{bmatrix} t\mathbf I_{D}+N^\text{com} \mathbf W^H \mathbf R_{\mathbf H}\mathbf W-\mathbf \Gamma & \sqrt{N^\text{com}}\mathbf W^H \\ \sqrt{N^\text{com}}\mathbf W & \mathbf R_{\mathbf H}^{-1} + \frac{1}{\sigma^2}\mathbf R_{\mathbf X} \end{bmatrix} \succeq \mathbf 0.
%\end{align}
The remaining issue is to deal with the nonconvex function $f(t,u) \triangleq tu$ involved in the objective function of (\ref{prob:JD_Xtu}), which can be handled by the MM method. Specifically, a proper convex surrogate function of $f(t,u)$ is defined by
\begin{align}
f(t,u;\varepsilon) \triangleq \frac{\varepsilon}{2}t^2 + \frac{1}{2\varepsilon}u^2,
\end{align}
where $f(t,u;\varepsilon)- f(t,u) = \frac{1}{2}(\sqrt{\varepsilon}t - \frac{1}{\sqrt{\varepsilon}}u)^2 \geq 0$ always holds for $\varepsilon > 0$. Particularly, when $\varepsilon = \frac{u}{t}$, we have
\begin{align}
f(t,u;\varepsilon)=&\ f(t,u),\nonumber \\
\nabla f(t,u;\varepsilon)= &\ \nabla f(t,u),
\end{align}
which satisfies the convergence conditions of constructing a surrogate function in the MM framework \cite[Section III-B]{MMAcce1}. Therefore, for the $i$-th iteration of the MM algorithm, we replace the nonconvex $f(t,u)$ with the convex $f(t,u;\varepsilon^{(i)})$ and iteratively solve the following convex problem:
\begin{align}\label{prob:JD_X_SCA}
\mathop \text{minimize}\limits_{\mathbf R_{\mathbf X}, \mathbf \Gamma,t,u} \quad &
\frac{\omega_1}{D} f(t,u;\varepsilon^{(i)}) \nonumber\\
& + \frac{\omega_2 }{M} \text{Tr}\left\{ \left(\mathbf R_{\mathbf G}^{-1} + \frac{1}{\sigma^2} \mathbf R_{\mathbf X} + \frac{L^\text{DT}}{\sigma^2} \mathbf W \mathbf W^H \right)^{-1} \right\} \nonumber \\
\text{subject to} \quad & \text{constraints of (\ref{prob:JD_Xtu})},
\end{align}
where $\varepsilon^{(i)} = \frac{u^{(i-1)}}{t^{(i-1)}}$ with $\{u^{(i-1)},t^{(i-1)}\}$ being the solutions to $\{u,t\}$ obtained in the $(i-1)$-th iteration.

By iteratively solving the convex problem in (\ref{prob:JD_X_SCA}) until convergence, we obtain a suboptimal solution of $\mathbf R_{\mathbf X}$. Moreover, according to the convergence analysis in \cite{MMAcce1}, this iterative procedure converges to a stationary point of problem (\ref{prob:JD_Xtu}).
In addition, by utilizing the analysis method in \cite{complexity}, solving problem (\ref{prob:JD_X_SCA}) requires a computational complexity of order $\mathcal O(\sqrt{M^2+N^2}(M^6+N^6))$.
After that, based on $\mathbf R_{\mathbf X}$, $\mathbf X$ can be recovered according to (\ref{decomp:R_X}).

\subsection{Optimization With Respect to the Transmit Beamforming}
We now consider optimization with respect to $\mathbf W$ with a fixed $\mathbf X$. Note that this subproblem has a similar form as $(\mathcal P_2)$, except that the fractional term in $\text{MSE}^\text{com}$ contained in the objective function of $(\mathcal P_2)$, i.e., $\frac{\mathbf W^H \mathbf{\hat H}^H \mathbf{\hat H} \mathbf W} {\text{Tr}\{ \mathbf W \mathbf W^H \mathbf R_{\mathbf \Delta} \} + \sigma^2}$, becomes $\frac{N^\text{com} \mathbf W^H \mathbf R_\mathbf{\hat H} \mathbf W} {\text{Tr}\{ \mathbf W \mathbf W^H \mathbf R_{\mathbf \Delta} \} + \sigma^2}$ here.
As a remedy, by defining $\mathbf{\tilde H} =\sqrt{N^\text{com}} \mathbf R_\mathbf{\hat H}^{1/2}$ and replacing $\mathbf{\hat H}$ with $\mathbf{\tilde H}$, the subproblem with respect to $\mathbf W$ has the same form as $(\mathcal P_2)$ and the proposed MM-based Algorithm~\ref{alg:MM} is also applicable here. The detailed process is omitted for brevity.

\begin{algorithm}[t]
\caption{Proposed Algorithm for Solving $(\mathcal P_3)$}
\label{alg:P3}
\begin{algorithmic}[1]
\STATE \textit{Initialization:} Set $i=0$ and initialize a feasible $ \mathbf W^{(0)}$.
\REPEAT
    \STATE Set $i=i + 1$.\\
    \STATE
    Iteratively solve the problem in (\ref{prob:JD_X_SCA}) until convergence and update $\mathbf X^{(i)}$ according to (\ref{decomp:R_X}). \\
    \STATE
    Update $\mathbf W^{(i)}$ utilizing Algorithm~\ref{alg:MM} by replacing $\mathbf{\hat H}$ with $\mathbf{\tilde H}$. \\
\UNTIL \textit{convergence}.
\end{algorithmic}
\end{algorithm}

To summarize, the complete procedure for solving $(\mathcal P_3)$ is listed in Algorithm~\ref{alg:P3} here.
It can be verified that the objective value of $(\mathcal P_3)$ is monotonically non-increasing over the alternating procedure and the solution set is compact, thus Algorithm~\ref{alg:P3} is guaranteed to converge.
In addition, the total computational complexity of Algorithm~\ref{alg:P3} is given by $\mathcal O( \mathcal I_{\text{AO}}(\mathcal I_{\mathbf X}\sqrt{M^2+N^2}(M^6+N^6) + \mathcal I_{\mathbf W}(M^3+N^3))$, where $\mathcal I_{\text{AO}}$, $\mathcal I_{\mathbf X}$, and $\mathcal I_{\mathbf W}$ denote the numbers of required iterations for the alternating procedure, for updating $\mathbf X$, and for updating $\mathbf W$, respectively.

\section{Structured Solutions Under Special Cases}
In this section, we provide structured and low-cost solutions to $(\mathcal P_1)$ and $(\mathcal P_3)$ under two different scenarios.
In the first scenario, we assume that the communication channel correlation $\mathbf R_{\mathbf H}$ and the radar TRM correlation $\mathbf R_{\mathbf G}$ have the same eigenvectors \cite{T.Naghibi2011}.
This assumption holds when $\mathbf R_{\mathbf H}$ or $\mathbf R_{\mathbf G}$ equals to identity matrix, corresponding to an uncorrelated channel in rich scattering environments.
In addition, this assumption also makes sense when $\mathbf H$ and $\mathbf G$ reuse the same scatterers in the vicinity of the ISAC transmitter.
We consider another commonly used least-squares (LS) criterion for estimating $\mathbf H$ and $\mathbf G$ in the second scenario.
In what follows, we investigate the structured solutions under the first scenario in Section~V-A and Section~V-B, and then discuss the results under the LS estimation in Section~V-C.

\subsection{Structured Solution for $(\mathcal P_1)$}
Let the eigenvalue decompositions (EVDs) of $\mathbf R_{\mathbf H} $ and $\mathbf R_{\mathbf G}$ be $\mathbf R_{\mathbf H} = \mathbf U_{\mathbf H} \mathbf \Lambda_{\mathbf H} \mathbf U_{\mathbf H}^H$ and $\mathbf R_{\mathbf G} = \mathbf U_{\mathbf G} \mathbf \Lambda_{\mathbf G} \mathbf U_{\mathbf G}^H$, respectively, where $\mathbf \Lambda_{\mathbf H} \triangleq \text{diag}\{[\lambda_{\mathbf H,1}, \cdots, \lambda_{\mathbf H,M}]^T \}$ and $\mathbf \Lambda_{\mathbf G} \triangleq \text{diag}\{[\lambda_{\mathbf G,1}, \cdots, \lambda_{\mathbf G,M}]^T \}$ are diagonal matrices including positive eigenvalues and $\mathbf U_{\mathbf H}$ and $\mathbf U_{\mathbf G}$ are the associated unitary matrices.
When $\mathbf U_{\mathbf H} = \mathbf U_{\mathbf G}= \mathbf U$, we have the following proposition for $(\mathcal P_1)$.
\begin{proposition}\label{prop:X_P1}
The optimal solution of $(\mathcal P_1)$ is $\mathbf X = \mathbf U \mathbf \Lambda_{\mathbf X}$, where $\mathbf \Lambda_{\mathbf X} = \left[ \text{diag}\{[\sqrt{x_1}, \cdots, \sqrt{x_{M}}]^T\}, \mathbf 0_{M \times (L^\text{CE} - M)} \right]$ is a rectangular diagonal matrix of size $M \times L^\text{CE}$. In particular, the optimal solution to $\{x_m \geq 0\}_{m=1}^M$, denoted by $\{x_m^\star\}_{m=1}^M$, satisfies
\begin{align}
\frac{\omega_1}{M\sigma^2} ( \lambda_{\mathbf H,m}^{-1} +&  x_m^\star/\sigma^2 )^{-2}
 +  \frac{\omega_2}{M\sigma^2} ( \lambda_{\mathbf G,m}^{-1} +  x_m^\star/\sigma^2 )^{-2}  \nonumber \\
 & =  \mu^\star,\ m \in \mathcal M \triangleq \{1,\cdots,M\}\label{xm_P1} \\
\sum_{m=1}^{M}  x_m^\star & = P^\text{CE},
\end{align}
where $\mu^\star \in (0,\bar \mu]$ and $\bar \mu\triangleq \text{min}(\hat\mu_1,\cdots,\hat\mu_M)$ with
\begin{align}\label{def:hat_mu}
\hat\mu_m = \frac{\omega_1}{M\sigma^2} \lambda_{\mathbf H,m}^{2} + \frac{\omega_2}{M\sigma^2}\lambda_{\mathbf G,m}^{2},\ m\in\mathcal M.
\end{align}
\end{proposition}
\begin{IEEEproof}
See Appendix~\ref{proof:X_P1}.
\end{IEEEproof}

Proposition~\ref{prop:X_P1} reveals a structured optimal solution of $(\mathcal P_1)$, from which we obtain that, the transmit directions of the training signal should be aligned with the eigenvectors of the correlations of the communication channel and the radar TRM. Moreover, based on this optimal structure, the original optimization with respect to $\mathbf X$ is converted to a lower-dimensional power allocation problem with respect to $\{x_m\}_{m=1}^M$, significantly reducing the computational complexity.
In particular, a numerical iterative method can be conducted to find $\{x_m^\star\}_{m=1}^M$ based on (\ref{xm_P1}).
Specifically, with a fixed $\mu^\star \in (0,\bar \mu]$, (\ref{xm_P1}) can be transformed into a quartic equation with respect to $x_m^\star$, whose real-valued positive root can be obtained analytically. Moreover, it is found from (\ref{xm_P1}) that $\sum_{m=1}^M x_m^\star$ is a decreasing function with respect to $\mu^\star$.
Thus, the optimal value of $\mu^\star$ can be obtained via a bisection search until $\sum_{m=1}^{M}  x_m^\star = P^\text{CE}$ holds.

\subsection{Structured Solution for $(\mathcal P_3)$}
As for $(\mathcal P_3)$, when $\mathbf U_{\mathbf H} = \mathbf U_{\mathbf G}= \mathbf U$, we can obtain the structure of its optimal solution and then propose a geometric programming (GP)-based iterative algorithm for the remaining power allocation problem.
Note that here we obtain the optimal solutions to $\mathbf X$ and $\mathbf W$ simultaneously instead of applying an alternating procedure as in Algorithm~\ref{alg:P3}.
To begin with, we give the following theorem concerning the optimal solution for $(\mathcal P_3)$.
\begin{theorem}\label{theorem:structureofP3}
The optimal solution of $(\mathcal P_3)$ satisfies the following structure:
\begin{align}\label{struture:p3}
\mathbf X = \mathbf U \mathbf \Lambda_{\mathbf X}, \quad \mathbf W = \mathbf U \mathbf \Lambda_{\mathbf W},
\end{align}
where $\mathbf \Lambda_{\mathbf W} = \left[ \text{diag}\{[\sqrt{w_1}, \cdots, \sqrt{w_{D}}]^T\}, \mathbf 0_{D \times (M - D)} \right]^T$ is a rectangular diagonal matrix of size $M \times D$.
\end{theorem}
\begin{IEEEproof}
See Appendix~\ref{proof:structureofP3}.
\end{IEEEproof}

Theorem~\ref{theorem:structureofP3} implies that, with the knowledge of channel statistics only, the optimal directions for both training and transmit beamforming are the same as those of the communication/sensing channels, so that the channel information can be efficiently exploited. This can be seen as an extension of the conclusions in MIMO communication systems \cite{jointdesign1,jointdesign2}.

Given the optimal structure in Theorem~\ref{theorem:structureofP3}, we transform $(\mathcal P_3)$ into a new optimization problem with respect to the power allocations of training and transmission signals, i.e., $\{x_m \geq 0\}_{m=1}^M$ and $ \{w_d \geq 0 \}_{d=1}^D$. By substituting (\ref{struture:p3}) into $(\mathcal P_3)$ and performing some algebraic operations, the reduced-dimensional problem is accordingly formulated as
\begin{align}\label{prob:JD_PA}
\mathop \text{minimize}\limits_{\{x_m \geq 0\}_{m=1}^M, \atop \{w_d \geq 0 \}_{d=1}^D} \quad &
\frac{\omega_1}{D}\! \sum_{d=1}^{D}\!\! \left( \!1\!\!+\!\! \frac{N^\text{com}}{\sum_{i=1}^D\!\!\frac{\lambda_{\mathbf H,i}\sigma^2w_i}{\lambda_{\mathbf H,i}x_i + \sigma^2} \!+\! \sigma^2}
\frac{\lambda_{\mathbf H,d}^2 w_d x_d}{ \lambda_{\mathbf H,d}x_d \!+\! \sigma^2 } \right)^{-1} \nonumber\\
&+ \frac{\omega_2}{M}
 \sum_{d=1}^{D} \left(\lambda_{\mathbf G,d}^{-1} + \frac{1}{\sigma^2 } x_d + \frac{L^\text{DT}}{\sigma^2 } w_d \right)^{-1} \nonumber\\
&+ \frac{\omega_2}{M}
 \sum_{m=D+1}^{M} \left(\lambda_{\mathbf G,m}^{-1} + \frac{1}{\sigma^2 } x_m\right)^{-1} \nonumber \\
\text{subject to} \quad & \sum_{m=1}^{M}x_m \leq P^\text{CE},\ \sum_{d=1}^{D}w_d \leq P^\text{DT}.
\end{align}
The term $\sum_{i=1}^D\frac{\lambda_{\mathbf H,i}\sigma^2w_i}{\lambda_{\mathbf H,i}x_i + \sigma^2}$ in the denominator of the communication MSE causes tight variable coupling.
To address this issue, we first transform (\ref{prob:JD_PA}) into the following equivalent problem by introducing real-valued auxiliary variables $ \xi_{d}, \kappa^\text{com}_{d}, \ d\in \mathcal D \triangleq\{1,\cdots,D\},$ $\kappa^\text{rad}_{m}, \ m \in \mathcal M,$ and $t$:
\begin{align}\label{prob:JD_PA_auxi}
\mathop \text{minimize}\limits_{\{x_m, \kappa_m^\text{rad}\}_{m=1}^M,t, \atop \{w_d , \xi_d, \kappa_d^\text{com} \}_{d=1}^D} \quad\!\! &
\frac{\omega_1}{D}  \sum_{d=1}^{D} \left( \kappa_d^\text{com} \right)^{-1}
+ \frac{\omega_2}{M} \sum_{m=1}^{M} \left(\kappa_m^\text{rad} \right)^{-1} \nonumber \\
\text{subject to} \quad\!\! & \sum_{m=1}^{M}x_m \leq P^\text{CE},\ \sum_{m=1}^{M}w_m \leq P^\text{DT},\nonumber\\
& \kappa_d^\text{com} \leq 1 +
\frac{N^\text{com} \lambda_{\mathbf H,d}^2 w_d x_d}{ t(\lambda_{\mathbf H,d}x_d +\sigma^2) },\ d \in \mathcal D \nonumber\\
& \kappa_m^\text{rad} \leq \lambda_{\mathbf G,m}^{-1} + \frac{1}{\sigma^2 } x_m + \frac{L^\text{DT}}{\sigma^2 } w_m,  \  m \in \mathcal D \nonumber\\
& \kappa_m^\text{rad} \leq \lambda_{\mathbf G,m}^{-1} + \frac{1}{\sigma^2 } x_m,  \  m \in \mathcal D\backslash \mathcal M \nonumber\\
& \xi_d \geq \frac{\lambda_{\mathbf H,d}\sigma^2w_d}{\lambda_{\mathbf H,d}x_d +\sigma^2 }, \ d \in \mathcal D \nonumber\\
& t \geq \sum_{d=1}^D \xi_{d} + \sigma^2.
\end{align}
The equivalence is proved by contradiction that the added inequality constraints regarding $\{ \kappa^\text{com}_{d}\}_{d=1}^D, $ $\{\kappa^\text{rad}_{m}\}_{m=1}^M,$ and $t$ in problem (\ref{prob:JD_PA_auxi}) must keep active at the optimality (see Appendix~\ref{proof:GP_equ} for details).
To proceed, we apply the idea of GP \cite[Section 4.5]{cvxbook} and define
\begin{align}\label{def:tilde}
t &= e^{\tilde t},\nonumber\\
x_m &= e^{\tilde x_m},\ \kappa_m^\text{rad} = e^{\tilde \kappa_m^\text{rad}},\ m \in \mathcal M \nonumber\\
w_d &= e^{\tilde w_d},\ \xi_d = e^{\tilde \xi_d},\ \kappa_d^\text{com} = e^{\tilde \kappa_d^\text{com}},\ d \in \mathcal D.
\end{align}
Substituting (\ref{def:tilde}) into problem (\ref{prob:JD_PA_auxi}) yields
\begin{align}\label{prob:GP}
\!\!\!\! \mathop \text{minimize}\limits_{\{\tilde x_m,\tilde \kappa_m^\text{rad}\}_{m=1}^M,\tilde t, \atop \{\tilde w_d ,\tilde \xi_d, \tilde \kappa_d^\text{com} \}_{d=1}^D} \  &
\frac{\omega_1}{D}  \sum_{d=1}^{D} e^{-\tilde \kappa_d^\text{com}} + \frac{\omega_2}{M}
 \sum_{m=1}^{M} e^{-\tilde \kappa_m^\text{rad}}\nonumber \\
\text{subject to} \quad & C_1: \sum_{m=1}^{M} e^{\tilde x_m} \leq L^\text{CE} P^\text{CE},\ \sum_{m=1}^{M}e^{\tilde w_m} \leq P^\text{DT},\nonumber\\
&C_2: \lambda_{\mathbf H,d}e^{\tilde x_d + \kappa_d^\text{com}+ \tilde t } \!+\! \sigma^2 e^{\kappa_d^\text{com}+\tilde t } \leq \sigma^2 e^{\tilde t}  \nonumber \\
& \quad\quad \!+\! \lambda_{\mathbf H,d} e^{\tilde x_d + \tilde t} +N^\text{com} \lambda_{\mathbf H,d}^2 e^{\tilde w_d} e^{\tilde x_d},\ d \in \mathcal D \nonumber\\
&C_3: e^{\tilde \kappa_m^\text{rad}} \leq \lambda_{\mathbf G,m}^{-1} \!+\! \frac{1}{\sigma^2 } e^{\tilde x_m} \!+\! \frac{L^\text{DT}}{\sigma^2 }e^{\tilde w_m}\!,  \ m \in \mathcal D \nonumber\\
&C_4: e^{\tilde \kappa_m^\text{rad}} \leq \lambda_{\mathbf G,m}^{-1} + \frac{1}{\sigma^2 } e^{\tilde x_m} ,  \ m \in \mathcal D\backslash \mathcal M\nonumber\\
&C_5: \lambda_{\mathbf H,d}\sigma^2 e^{\tilde w_d - \tilde \xi_d} \leq \lambda_{\mathbf H,d}e^{\tilde x_d} +\sigma^2 , \ d \in \mathcal D \nonumber\\
&C_6: \sum_{d=1}^D e^{\tilde \xi_d - \tilde t} + \sigma^2 e^{- \tilde t}\leq 1.
\end{align}
It is seen that most monomial functions in problem (\ref{prob:JD_PA_auxi}) is now converted into convex forms in problem (\ref{prob:GP}), except the nonconvex constraints $C_2 -C_5$. To address this issue, we employ the successive convex approximation (SCA) technique \cite{SCAconvergence} by applying the following first-order Taylor expansion on the constraints $C_2 -C_5$:
\begin{align}\label{inequ:sca}
e^{z} \geq e^{z^{(i-1)}} + e^{z^{(i-1)}} \left( z - z^{(i-1)} \right) \triangleq \underline f\left(z;z^{(i-1)}\right),
\end{align}
where $z^{(i-1)}$ denotes the solution in the $(i-1)$-th iteration of the SCA algorithm. As a result, we obtain a convex approximate problem in the $i$-th iteration, given by
\begin{align}\label{prob:GP_SCA}
\mathop \text{minimize}\limits_{\{\tilde x_m,\tilde \kappa_m^\text{rad}\}_{m=1}^M,\tilde t, \atop \{\tilde w_d ,\tilde \xi_d, \tilde \kappa_d^\text{com} \}_{d=1}^D} \ &
\frac{\omega_1}{D} \sum_{d=1}^{D} e^{\tilde t -\tilde \kappa_d^\text{com}} + \frac{\omega_2}{M}
 \sum_{m=1}^{M} e^{-\tilde \kappa_m^\text{rad}}\nonumber \\
\text{subject to}\quad &  C_1, C_2',C_3',C_4',C_5' , C_6,
\end{align}
where $C_2' -C_5'$ are, respectively, given by
\begin{align}\label{def:C'}
&C_2'\!:\! \lambda_{\mathbf H,d}e^{\tilde x_d + \kappa_d^\text{com} + \tilde t} \!\!+\! \sigma^2 e^{\kappa_d^\text{com}+ \tilde t}  \!\!\leq\!  \sigma^2 \underline \lambda_{\mathbf H,d} \underline f(\tilde t \!+\! \tilde x_d;\tilde t^{(i-1)} \!+\! \tilde x_{d}^{(i-1)}) \nonumber \\
& \quad\quad \!\!\!+\!\! f(\tilde t;\!\tilde t^{(i-1)}\!)  \!\!+\!\! N^\text{com} \lambda_{\mathbf H,d}^2 \underline f(\tilde x_d \!+\! \tilde w_d;\!\tilde x_{d}^{(i-1)\!} \!\!+\!\! \tilde w_{d}^{(i-1)}\!)
,\  d \!\in\! \mathcal D\nonumber\\
&C_3'\!:\! e^{\tilde \kappa_m^\text{rad}} \!\!\leq\!\!
\lambda_{\mathbf G,m}^{-1} \!\! +\!\! \frac{1}{\sigma^2 } \underline f(\tilde x_m;\!\tilde x_{m}^{(i-1)}\!) \!+ \!\! \frac{L^\text{DT}}{\sigma^2 } \underline f(\tilde w_m;\!\tilde w_{m}^{(i-1)}\!),  m \!\in\! \mathcal D \nonumber\\
&C_4'\!:\! e^{\tilde \kappa_m^\text{rad}} \leq
\lambda_{\mathbf G,m}^{-1} + \frac{1}{\sigma^2 } \underline f(\tilde x_m;\tilde x_{m}^{(i-1)}),  \  m \in \mathcal D\backslash \mathcal M \nonumber\\
&C_5'\!:\! \lambda_{\mathbf H,d}\sigma^2 e^{\tilde w_d - \tilde \xi_d} \leq  \sigma^2 \!+\! \lambda_{\mathbf H,d} \underline f(\tilde x_d;\tilde x_{d}^{(i-1)}), \  d \in \mathcal D.
\end{align}
Since problem (\ref{prob:GP_SCA}) is now of the convex form, it can be readily solved via the CVX toolbox \cite{CVXtool}.

\begin{algorithm}[t]
\caption{Proposed Algorithm for Solving problem (\ref{prob:JD_PA})}
\label{alg:SCA}
\begin{algorithmic}[1]
\STATE \textit{Initialization:} Set $i=0$, initialize the feasible $\{x_m^{(0)}\}_{m=1}^M$ and $ \{w_d^{(0)} \}_{d=1}^D$, and $t^{(0)} =  \sum_{d=1}^D \frac{\lambda_{\mathbf H,d}\sigma^2w_d^{(0)}}{\lambda_{\mathbf H,d}x_d^{(0)} +\sigma^2 } + \sigma^2$.
\STATE Calculate $\tilde x_m^{(0)} = \log( x_m^{(0)}),\ m\in\mathcal M$, $\tilde w_d^{(0)} = \log( w_d^{(0)}),\ d \in\mathcal D$, and $\tilde t^{(0)} = \log( t^{(0)})$.
\REPEAT
    \STATE Set $i=i + 1$.\\
    \STATE
    Solve the problem in (\ref{prob:GP_SCA}) with $ \tilde x_m^{(i-1)},\ m\in\mathcal M$, $ \tilde w_d^{(i-1)},\ d\in\mathcal D$, and $\tilde t^{(i-1)}$.
    \STATE Update $\{\tilde x_m^{(i)}\}_{m=1}^M$ and $ \{ \tilde w_d^{(i)} \}_{d=1}^D$, and $\tilde t^{(i)}$. \\
\UNTIL \textit{convergence}.
\end{algorithmic}
\end{algorithm}

To summarize, we list the procedure for solving problem (\ref{prob:JD_PA}) in Algorithm~\ref{alg:SCA}.
By iteratively solving (\ref{prob:GP_SCA}), a sequence of solutions can be successively generated which converge
to a KKT point of problem (\ref{prob:JD_PA}) \cite{SCAconvergence}. Finally, substituting the obtained $\{x_m\}_{m=1}^M$ and $ \{w_d \}_{d=1}^D$ into (\ref{struture:p3}), we obtain the optimal training and transmission signals for the joint design problem in $(\mathcal P_3)$. Moreover, according to \cite{GPcomplexity}, solving the convex GP in (\ref{prob:GP_SCA}) has a complexity $\mathcal O((D+M)^{3.5})$, which is much lower than that of Algorithm~\ref{alg:P3}.

\subsection{Solutions for LS Estimation}
When considering the LS criterion, the MSEs for estimating $\mathbf H$ and $\mathbf G$ become $\text{MSE}^\text{CE} = \sigma^2\text{Tr}\left\{ (\mathbf X\mathbf X^H)^{-1} \right\}$ and $\text{MSE}^\text{rad}= \sigma^2 \text{Tr}\left\{( \mathbf X\mathbf X^H + L^\text{DT}\mathbf W \mathbf W^H)^{-1} \right\}$, respectively.
In this case, it is easily seen that $(\mathcal P_1)$ has a closed-form optimal solution as $\mathbf X\mathbf X^H = \frac{P^\text{CE}}{M} \mathbf I_M$. As for $(\mathcal P_3)$, the optimal structures proposed in Theorem~\ref{theorem:structureofP3} are still applicable and a similar procedure can be conducted to solve the remaining power allocation problem.

\section{Extension to MI-Based Design}

\begin{table*}[t]
\caption{MSE-Based and MI-Based Performance Metrics for the Considered MIMO ISAC System}
\label{table:metric}
    \centering
	\begin{tabular}{|c|c|c|c|}
		\hline
\multicolumn{2}{|c|}{\textbf{Task}} & \textbf{MSE-based metric} & \textbf{MI-based metric} \\
        \hline
        &Channel estimation
        & $\text{MSE}^\text{CE}  = \text{Tr}\left\{  \left(\mathbf R_{\mathbf H}^{-1} + \frac{1}{\sigma^2} \mathbf X\mathbf X^H\right)^{-1}  \right\}$
        & $ \text{MI}^\text{CE} = \log\det \left( \mathbf I_{M} + \frac{1}{\sigma^2} \mathbf R_\mathbf H \mathbf X \mathbf X^H  \right)$\\
        \cline{2-4}
        %\multirow{1}{*}{com}
        \textbf{com}
        & \multirow{4}{*}{Data transmission}
        & $\text{MSE}^\text{com}
= \text{Tr}\left\{ \left( \mathbf I_{D} +  \frac{\mathbf W^H \mathbf{\hat H}^H \mathbf{\hat H} \mathbf W} {\text{Tr}\{ \mathbf W \mathbf W^H \mathbf R_{\mathbf \Delta} \} + \sigma^2}  \right)^{-1}\right\}$
        & $\text{MI}^\text{com}
= \log\det \left( \mathbf I_{D} +  \frac{\mathbf W^H \mathbf{\hat H}^H \mathbf{\hat H} \mathbf W} {\text{Tr}\{ \mathbf W \mathbf W^H \mathbf R_{\mathbf \Delta} \} + \sigma^2}  \right)$ \\

        &
        & $\overline{\text{MSE}}^\text{com} = \text{Tr}\left\{ \left( \mathbf I_{D} +  \frac{N^\text{com} \mathbf W^H \mathbf R_\mathbf{\hat H} \mathbf W} {\text{Tr}\{ \mathbf W \mathbf W^H \mathbf R_{\mathbf \Delta} \} + \sigma^2} \right)^{-1}\right\}$
        &  $\overline{\text{MI}}^\text{com} = \log\det \left( \mathbf I_{D} +  \frac{N^\text{com} \mathbf W^H \mathbf R_\mathbf{\hat H} \mathbf W} {\text{Tr}\{ \mathbf W \mathbf W^H \mathbf R_{\mathbf \Delta} \} + \sigma^2}  \right)$ \\
        \hline \textbf{rad}
        &TRM estimation
        & $\text{MSE}^\text{rad} = \text{Tr}\left\{ \left(\mathbf R_{\mathbf G}^{-1} + \frac{1}{\sigma^2} (\mathbf X\mathbf X^H + L^\text{DT}\mathbf W \mathbf W^H) \right)^{-1} \right\}$
        & $\text{MI}^\text{rad} = \log\det \left( \mathbf I_{M} + \frac{1}{\sigma^2} \mathbf R_\mathbf G  (\mathbf X \mathbf X^H + L_\text{DT} \mathbf W \mathbf W^H) \right)$ \\
        \hline
	\end{tabular}%
\end{table*}

Apart from the MSE minimization, maximizing the MI is another popular design criterion in MIMO systems.
In this section, we extend the previous MSE-based ISAC training and transmission optimization to the MI maximization criterion.
Specifically, we derive the MI-based performance metrics for the considered ISAC system following \cite{B.TangTSP2019,Y.Yang2007,C.Xingjointdesign}, where the details are omitted due to the page limit. We present a summary and comparison of the MSE-based and MI-based metrics for the considered training and transmission design in Table~\ref{table:metric}. It is worth noting that, owing to the similar forms and inherent relationship between MSE and MI metrics, the MI-based training and transmission optimization can be formulated and addressed in a similar way by utilizing the approaches proposed previously for the MSE-based design (with some minor modifications). Further details are provided as follows.

\subsection{Instantaneous CSI-Based Sequential Design}

\subsubsection{ISAC Training Design}
The optimization of the ISAC training signal $\mathbf X$ is formulated as
\begin{align}
(\mathcal P_1'):\
 \mathop \text{maximize}\limits_{\mathbf X} \quad &
\frac{\omega_1}{M}  \log\det \left( \mathbf I_{M} +  \mathbf R_\mathbf H \mathbf X \mathbf X^H/\sigma^2  \right) \nonumber\\
& + \frac{\omega_2}{M} \log\det \left( \mathbf I_{M} + \mathbf R_\mathbf G \mathbf X \mathbf X^H /\sigma^2  \right) \nonumber \\
\text{subject to} \quad & \text{Tr}\left\{ \mathbf X \mathbf X^H\right\} \leq P^\text{CE},
\end{align}
where the MIs are normalized by the number of transmit antennas $M$.
This problem is convex with respect to $\mathbf R_{\mathbf X} \triangleq \mathbf X \mathbf X^H$ and can be readily solved. Then, the optimal $\mathbf X$ can be recovered according to (\ref{decomp:R_X}).

\subsubsection{ISAC Robust Transmission Design}
The optimization of the transmit beamforming $\mathbf W$ is formulated as:
\begin{align}
(\mathcal P_2'):\
\mathop \text{maximize}\limits_{\mathbf W} \quad &
\frac{\omega_1}{D} \log\det \left( \mathbf I_{D} +  \frac{\mathbf W^H \mathbf{\hat H}^H \mathbf{\hat H} \mathbf W} {\text{Tr}\{ \mathbf W \mathbf W^H \mathbf R_{\mathbf \Delta} \} + \sigma^2}  \right) \nonumber\\
&\!\!\!\!\!\!\!\!\!\!\!\!\!\!\!\!\!\!\!\!\!\!\!\!\!\!\!\!\!\!\!\! +  \frac{\omega_2}{M} \log\det \left( \mathbf I_{M} + \mathbf R_\mathbf G  (\mathbf X^\star (\mathbf X^\star)^H + L_\text{DT} \mathbf W \mathbf W^H)/\sigma^2 \right) \nonumber \\
\text{subject to} \quad & \text{Tr}\left\{ \mathbf W \mathbf W^H\right\} \leq P^\text{DT},
\end{align}
where $\mathbf X^\star$ represents the solution of $(\mathcal P_1')$.
This nonconvex and challenging problem can be solved employing the MM algorithm.
Concretely, by following a similar procedure as in Appendix~\ref{proof:MM_P2} to construct the surrogate function, we can obtain an identical problem as (\ref{prob:p2MM}) and then solve it iteratively. The difference lies in that $\lambda$, $\mathbf \Psi$, and $ \mathbf \Pi^H$ in (\ref{prob:p2MM}) become $\lambda' = \lambda_\text{max} (\mathbf \Xi_0^{-1} \mathbf W^H_0  \mathbf R_{\mathbf G'}^{-1} \mathbf A_0^{-1}\mathbf R_{\mathbf G'}^{-1} \mathbf W_0 \mathbf \Xi_0^{-1} ) \lambda_\text{max} ( \mathbf R_{\mathbf G'}^{-1}),
\mathbf \Psi' = \mathbf{\hat H}^H \mathbf Q_0^{-1} \mathbf{\hat H} \mathbf W_0 \mathbf B_0^{-1} \mathbf W^H_0 \mathbf{\hat H}^H \mathbf Q_0^{-1}\mathbf{\hat H} + \text{Tr}\{ \mathbf Q_0^{-1} \mathbf{\hat H} \mathbf W_0 \mathbf B_0^{-1} \mathbf W^H_0 \\ \mathbf{\hat H}^H  \mathbf Q_0^{-1}\}   \mathbf R_{\mathbf \Delta}$,
and
$\mathbf \Pi'^H = \frac{\omega_1}{D} \mathbf B_0^{-1} \mathbf W^H_0 \mathbf{\hat H}^H \mathbf Q^{-1}_0 \mathbf{\hat H} +
\frac{\omega_2}{M} (\mathbf \Xi_0^{-1}\mathbf W_0^H \mathbf R_{\mathbf G'}^{-1} \mathbf A_0^{-1}\mathbf R_{\mathbf G'}^{-1}  +  \lambda' \mathbf W_0^H
- \mathbf \Xi_0^{-1} \mathbf W^H_0   \mathbf R_{\mathbf G'}^{-1} \mathbf A_0^{-1}\mathbf R_{\mathbf G'}^{-1} \\ \mathbf W_0 \mathbf \Xi_0^{-1} \mathbf W_0^H  \mathbf R_{\mathbf G'}^{-1} )$ here, respectively, in which
$\mathbf A_0 \triangleq \mathbf R_{\mathbf G'}^{-1} -   \mathbf R_{\mathbf G'}^{-1} \mathbf W_0 \mathbf \Xi_0^{-1}  \mathbf W_0^H  \mathbf R_{\mathbf G'}^{-1}$ and
$\mathbf B_0 \triangleq  \mathbf I_{D} -  \mathbf W_0^H \mathbf{\hat H}^H \mathbf Q_0^{-1} \mathbf{\hat H} \mathbf W_0$.

\subsection{Statistical CSI-Based Joint Design}
The joint optimization of $\mathbf X$ and $\mathbf W$ is formulated as:
\begin{align}
(\mathcal P_3'):\
\mathop \text{maximize}\limits_{\mathbf X,\mathbf W} \quad &
\frac{\omega_1}{D} \log\det \left( \mathbf I_{D} +  \frac{N^\text{com} \mathbf W^H \mathbf R_\mathbf{\hat H} \mathbf W} {\text{Tr}\{ \mathbf W \mathbf W^H \mathbf R_{\mathbf \Delta} \} + \sigma^2}  \right) \nonumber\\
&\!\!\!\!\!\!\!\!\!\!\!\!\!\!\!\!\!\!\!\!\!\!\!\! + \frac{\omega_2}{M} \log\det \left( \mathbf I_{M} + \mathbf R_\mathbf G  (\mathbf X \mathbf X^H + L_\text{DT} \mathbf W \mathbf W^H)/\sigma^2 \right) \nonumber \\
\text{subject to} \quad
& \text{Tr}\left\{ \mathbf X \mathbf X^H\right\} \leq P^\text{CE},\
 \text{Tr}\left\{ \mathbf W \mathbf W^H\right\} \leq P^\text{DT},
\end{align}
which can be solved via the AO framework. Concretely, by introducing a series of auxiliary variables and following a similar procedure as in Appendix~\ref{proof:prob_equ}, the subproblem with respect to $\mathbf X$ can be equivalently transformed into the following form:
\begin{align}\label{prob:R2_MI_X}
\mathop \text{maximize}\limits_{\mathbf R_{\mathbf X},\mathbf \Gamma,t \geq 0, u\geq 0  } \quad &
\frac{\omega_1}{D} \frac{u^2}{t}
+ \frac{\omega_2}{M} \log\det ( \mathbf I_{M} + \mathbf R_\mathbf G \mathbf R_\mathbf X/\sigma^2 \nonumber \\
&\quad\quad\quad\quad\quad\quad\quad\quad\quad + L_\text{DT} \mathbf R_\mathbf G  \mathbf W \mathbf W^H/\sigma^2 ) \nonumber \\
\text{subject to} \quad &  \text{constraints of problem (\ref{prob:R2_t})}, \nonumber \\
& u^2 \leq \log\det \left(  t \mathbf I_{D} + N^\text{com} \mathbf W^H (\mathbf R_{\mathbf H} - \mathbf \Gamma) \mathbf W \right).
\end{align}
The nonconvexity lies in the function $\frac{u^2}{t}$ in the objective function. Since $\frac{u^2}{t}$ is a joint convex function with respect to $\{u,t\}$, we can solve this problem based on the SCA framework. In particular, by exploiting the first-order Taylor expansion: $\frac{u^2}{t} \geq 2 \frac{ u^{(i-1)}}{ t^{(i-1)} } u - \left(\frac{ u^{(i-1)}}{ t^{(i-1)} } \right)^2 t \triangleq \underline g(t,u;t^{(i-1)}, u^{(i-1)})$, in each iteration we solve the following convex problem:
\begin{align}
\mathop \text{maximize}\limits_{\mathbf R_{\mathbf X},\mathbf \Gamma,t \geq 0, u\geq 0  } \quad &
\frac{\omega_1}{D} g\left(t,u;t^{(i-1)}, u^{(i-1)} \right) \nonumber \\
& \!\!\!\!\!\!\!\!\!\!\!\!\!\!\!\!\!\!\!\!\!\!\!+ \frac{\omega_2}{M} \log\det ( \mathbf I_{M} + \mathbf R_\mathbf G \mathbf R_\mathbf X/\sigma^2+ L_\text{DT} \mathbf R_\mathbf G  \mathbf W \mathbf W^H /\sigma^2 ) \nonumber \\
\text{subject to} \quad &  \text{constraints of problem (\ref{prob:R2_MI_X})}.
\end{align}
On the other hand, by defining $\mathbf{\tilde H} =\sqrt{N^\text{com}} \mathbf R_\mathbf{\hat H}^{1/2}$ and replacing $\mathbf{\hat H}$ with $\mathbf{\tilde H}$, the subproblem with respect to $\mathbf W$ can be addressed using the MM algorithm proposed for $(\mathcal P_2')$.

\subsection{Low-Cost Solution Under Special Case}
We discuss the low-cost solution for the challenging problem $(\mathcal P_3')$.
For the considered special case where $\mathbf R_{\mathbf H}$ and $\mathbf R_{\mathbf G}$ have the same eigenvectors, it can be similarly verified that the structures obtained in Theorem~\ref{theorem:structureofP3}, i.e., $\mathbf X = \mathbf U \mathbf \Lambda_{\mathbf X}$ and $  \mathbf W = \mathbf U \mathbf \Lambda_{\mathbf W}$, remain optimal.
Substituting these structures and following a similar procedure as shown in Appendix~\ref{proof:GP_equ}, we finally transform $(\mathcal P_3')$ into the following equivalent problem:
\begin{align}\label{prob:R2_MI_xi}
\mathop \text{maximize}\limits_{\{x_m\}_{m=1}^M,t, \atop \{w_d ,\xi_{d} \}_{d=1}^D} \quad &
\frac{\omega_1}{D}
\sum_{d=1}^{D}\log\left( \kappa_d^\text{com} \right)
+ \frac{\omega_2}{M}
 \sum_{m=1}^{M} \log \left( \kappa_m^\text{rad} \right) \nonumber \\
\text{subject to}
\quad &  \text{constraints of problem (\ref{prob:JD_PA_auxi})},
\end{align}
where the left-hand sides of the third and fourth constraints are replaced with $\lambda_{\mathbf G,m}^{-1}\kappa_m^\text{rad} $.
%The equivalence is proved by contradiction that the added inequality constraints regarding $\{ \kappa^\text{com}_{d}\}_{d=1}^D, $ $\{\kappa^\text{rad}_{m}\}_{m=1}^M,$ and $t$ in problem (\ref{prob:R2_MI_xi}) must keep active at the optimality.
To proceed, we substitute (\ref{def:tilde}) and utilize the SCA framework to solve problem (\ref{prob:R2_MI_xi}). In particular, the convex approximation problem solved in the $i$-th iteration is given by
\begin{align}
\mathop \text{maximize}\limits_{\{\tilde x_m,\tilde \kappa_m^\text{rad}\}_{m=1}^M,\tilde t, \atop \{\tilde w_d ,\tilde \xi_d, \tilde \kappa_d^\text{com} \}_{d=1}^D} \quad &
\frac{\omega_1}{D} \sum_{d=1}^{D} \tilde \kappa_m^\text{rad} + \frac{\omega_2}{M}
 \sum_{m=1}^{M} \tilde \kappa_m^\text{com} \nonumber \\
\text{subject to}\quad &  C_1, C_2',C_3',C_4',C_5' , C_6,
\end{align}
where the left-hand sides of $C_3'$ and $C_4'$ are replaced with $\lambda_{\mathbf G,m}^{-1} e^{\tilde \kappa_m^\text{rad}} $.

\begin{figure}[t]
\begin{center}
      \epsfxsize=7.0in\includegraphics[scale=0.52]{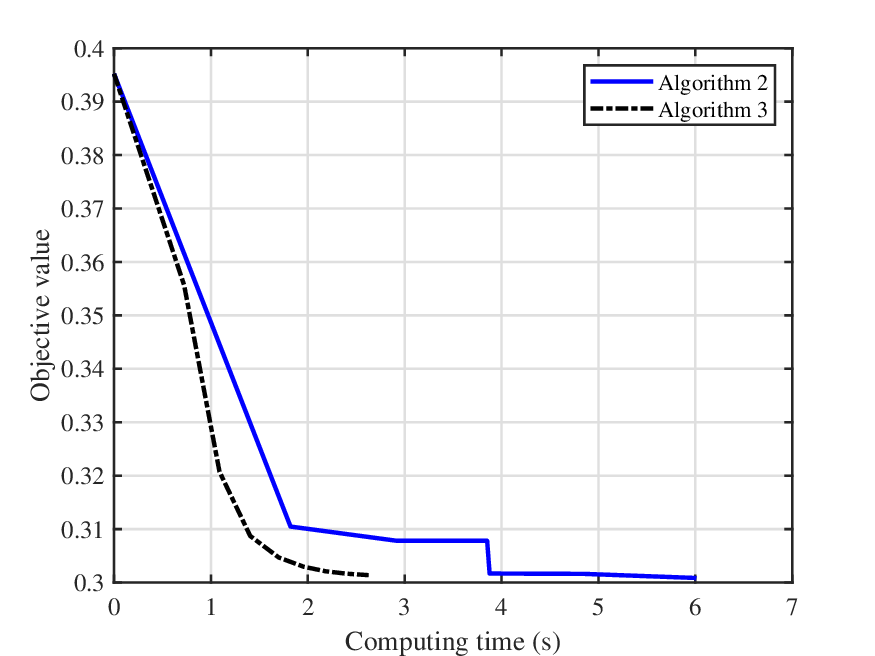}
      \caption{Objective value versus computing time for Algorithms 2 and 3.}\label{fig:iteration}
    \end{center}
\end{figure}

\begin{figure}[t]
\begin{center}
      \epsfxsize=7.0in\includegraphics[scale=0.52]{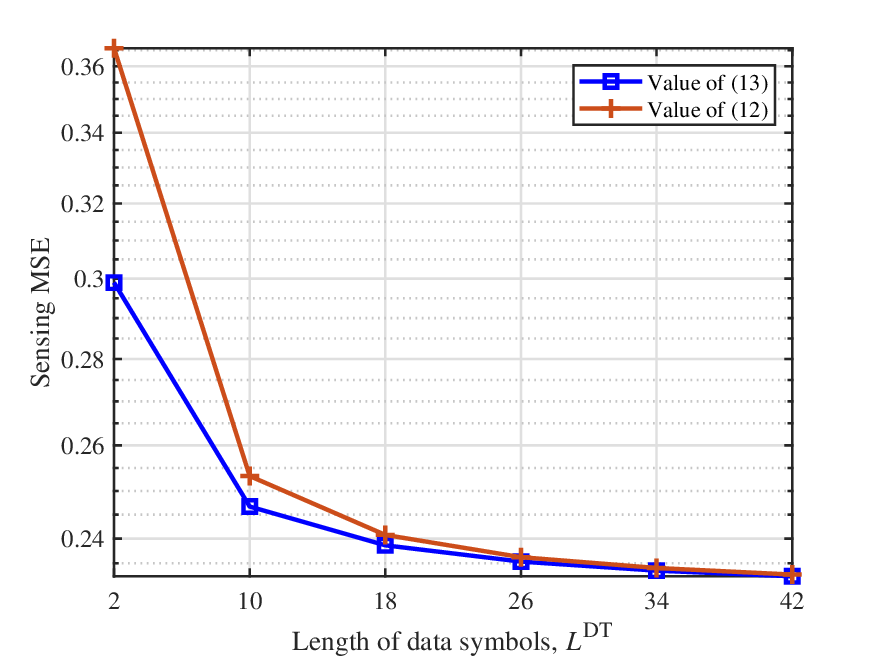}
      \caption{Accurate and approximated sensing MSEs comparison versus the length of data symbol $L^\text{DT}$.}\label{fig:verify_SS}
    \end{center}
\end{figure}

%\vspace{-10pt}
\section{Simulation Results}\label{section:simulation}
In this part, we evaluate the performances of the proposed design schemes. The antenna numbers of the ISAC transmitter, the communication receiver, and the radar receiver are set to $M = 8$, $N^\text{com} = 4$, and $N^\text{rad} = 8$ respectively. The number of data streams for MIMO communication is set to $D = 4$. Assume that a channel fading block contains $L = 40$ time slots, in which the training stage occupies $L^\text{CE} = M = 8$ slots while the remaining $L^\text{DT} =  L - L^\text{CE} = 32$ slots are left for data transmission.
The SNRs of training and data transmission are defined by $\gamma^\text{CE} = P^\text{CE}/(L^\text{CE}\sigma^2)$ and $\gamma^\text{DT} = P^\text{DT}/\sigma^2$, respectively. For simplicity, the noise is normalized as $\sigma^2 = 1$ and the SNRs are adjusted by tuning the corresponding transmit powers $P^\text{CE}$ and $P^\text{DT}$. The weighting factors for communication and sensing are set to $\omega_1 = \omega_2 = 0.5$ for fairness, unless stated otherwise.
We employ the commonly used exponential correlation model \cite{Channelcorrelation} for the correlation matrices $\mathbf R_{\mathbf H}$ and $\mathbf R_{\mathbf G}$. Concretely, $\mathbf R_{\mathbf H}$ is modeled as $[\mathbf R_{\mathbf H}]_{i,j} = \rho^{|i-j | },\ \forall i,j$, where $[\mathbf R_{\mathbf H}]_{i,j}$ denotes the $(i,j)$-th element of $\mathbf R_{\mathbf H}$ and $0 \leq \rho < 1$ stands for the coefficient of spatial correlation. We set $\rho = 0.5$ for $\mathbf R_{\mathbf H}$. The correlation matrix of $\mathbf G$ is set as $\mathbf R_{\mathbf G} = \mathbf I_M$.
All the MSE results are averaged over 1000 independent channel realizations based on the Monte Carlo method. Specifically, given $\mathbf R_{\mathbf H}$ and $\mathbf R_{\mathbf G}$, the channel realizations $\mathbf H$ and $\mathbf G$ are generated as $\mathbf H = \mathbf Z_{\mathbf H} \mathbf R_{\mathbf H}^{1/2}$ and $\mathbf G = \mathbf Z_{\mathbf G} \mathbf R_{\mathbf G}^{1/2}$, respectively, where the entries of $\mathbf Z_{\mathbf H}$ and $\mathbf Z_{\mathbf G}$ are i.i.d. complex Gaussian random variables with zero mean and unit variance \cite{J.YaoTWC}.

We first compare the convergence behaviours of the developed iterative Algorithms 2 and 3 in Fig.~\ref{fig:iteration}. We observe that the proposed Algorithm~2 and Algorithm~3 typically converge within several seconds and attain the same objective value.
Moreover, the advantage of Algorithm~3 in terms of the computational rate can be clearly observed, which verifies that Algorithm~3 is an effective low-cost alternative of Algorithm~2 under the special case.

Then, we validate the accuracy of the approximation in (\ref{MSE:rad}). Specifically, we independently generate the elements of $\mathbf S$ with quadrature phase shift keying (QPSK) symbols and calculate (\ref{MSE:rad_real}) and (\ref{MSE:rad}), with $\mathbf X$ and $\mathbf W$ determined using the method proposed in Section III. The comparison between the values of (\ref{MSE:rad_real}) and (\ref{MSE:rad}) versus different values of $L^\text{DT}$ is presented in Fig.~\ref{fig:verify_SS}.
From the figure, it is evident that the error between (\ref{MSE:rad_real}) and (\ref{MSE:rad}) decreases as $L^\text{DT}$ increases, and the approximation mismatch is negligible when $L^\text{DT} \geq 26$.

\begin{figure}[t]
\centering
\subfigure[MSE of data transmission.]{
\begin{minipage}[t]{0.9\linewidth}
\centering
\includegraphics[scale=0.52]{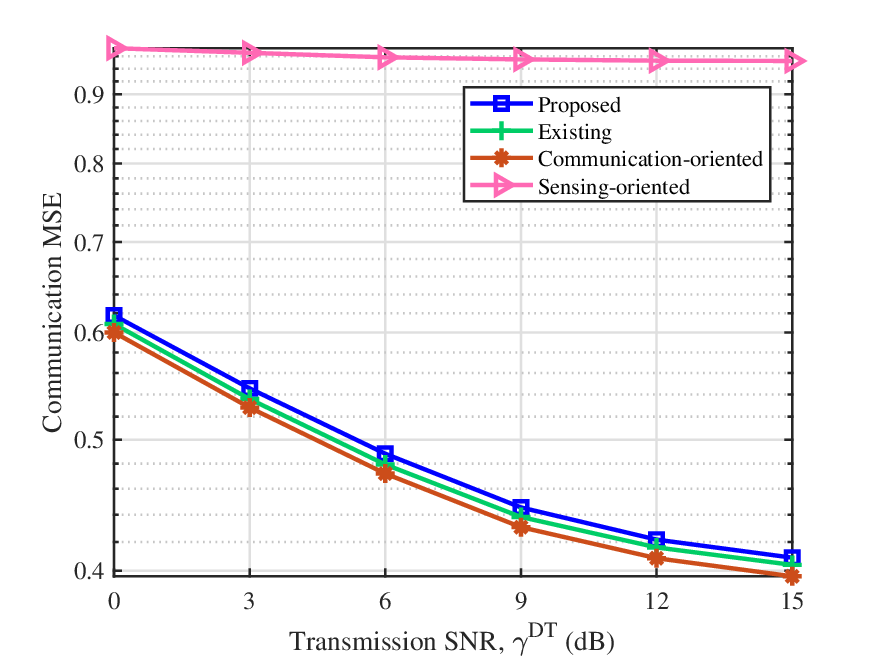}
%\caption{fig1}
\end{minipage}%
}%
\\
\subfigure[MSE of radar TRM estimation.]{
\begin{minipage}[t]{0.9\linewidth}
\centering
\includegraphics[scale=0.52]{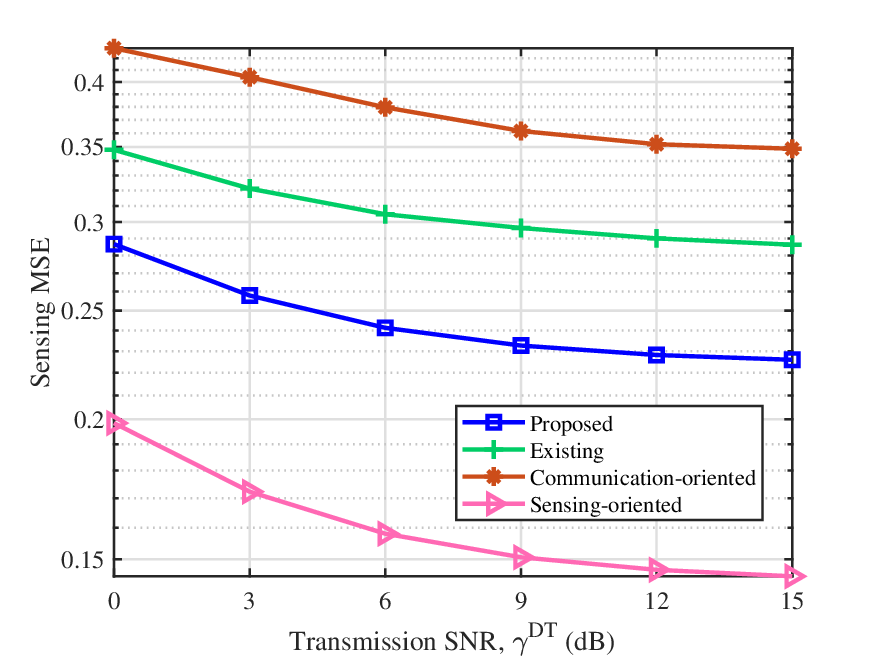}
%\caption{fig2}
\end{minipage}%
}%
\centering
\caption{MSE performances versus $\gamma^\text{DT}$ with instantaneous CSI feedback.}\label{fig:ins_PCE}
\end{figure}

To proceed, we introduce the following benchmark schemes for performance comparisons.

\textbf{1) Existing ISAC Design Scheme:}
The training signal $\mathbf X$ is optimized to minimize $\text{MSE}^\text{CE}$ solely, denoted by $\mathbf X^\star$. However, it is also received at the radar receiver and utilized for sensing purpose, as in \cite{Spatio-temporal}. The ISAC beamforming matrix $\mathbf W$ is optimized through
\begin{align}\label{bench1}
\mathop \text{minimize}\limits_{\mathbf W} \quad &
\frac{\omega_1}{D} \text{Tr}\left\{ \left( \mathbf I_{D} +  \frac{\mathbf W^H \mathbf{\hat H}^H \mathbf{\hat H} \mathbf W} {\text{Tr}\{ \mathbf W \mathbf W^H \mathbf R_{\mathbf \Delta} \} + \sigma^2} \right)^{-1}\right\} \nonumber\\
&\!\!\!\!\!\!\!\!\!\!\!\!\!\! +\frac{\omega_2 }{M } \text{Tr}\left\{ \left(\mathbf R_{\mathbf G}^{-1}  \!+\! \mathbf X^\star (\mathbf X^\star)^H/\sigma^2 \!+\! L^\text{DT} \mathbf W \mathbf W^H/ \sigma^2\right)^{-1} \right\} \nonumber\\
\text{subject to} \quad & \text{Tr}\left\{ \mathbf W \mathbf W^H\right\} \leq P^\text{DT}.
\end{align}

\textbf{2) Communication-Oriented Scheme:} Omit the sensing requirements, i.e., letting $\omega_2 = 0$, when solving $(\mathcal P_1)-(\mathcal P_3)$.

\textbf{3) Sensing-Oriented Scheme:} Jointly design $\mathbf X$ and $\mathbf W$ to minimize $\text{MSE}^\text{rad}$.

\begin{figure}[t]
\centering
\subfigure[MSE of data transmission.]{
\begin{minipage}[t]{0.9\linewidth}
\centering
\includegraphics[scale=0.52]{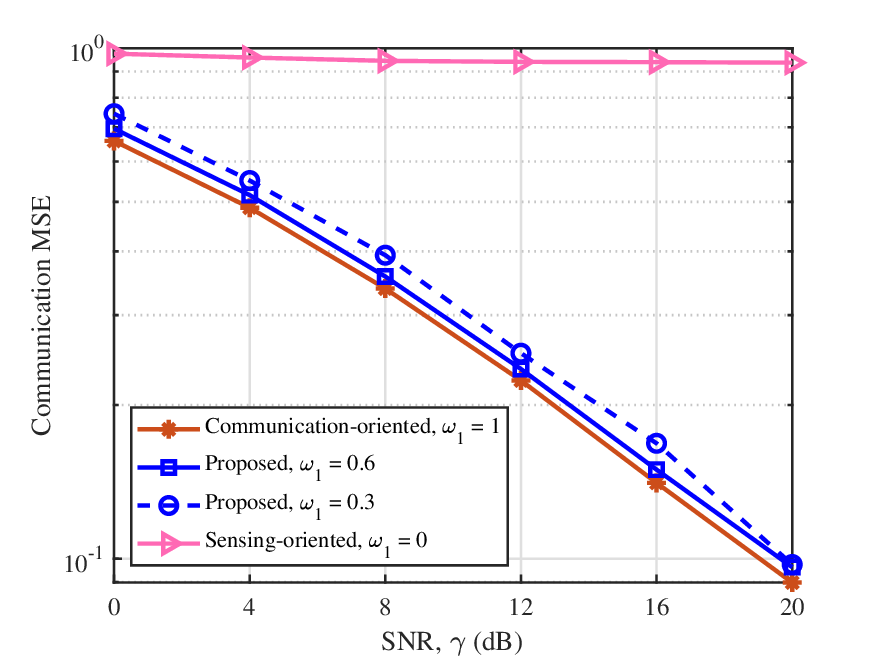}
%\caption{fig1}
\end{minipage}%
}%
\\
\subfigure[MSE of radar TRM estimation.]{
\begin{minipage}[t]{0.9\linewidth}
\centering
\includegraphics[scale=0.52]{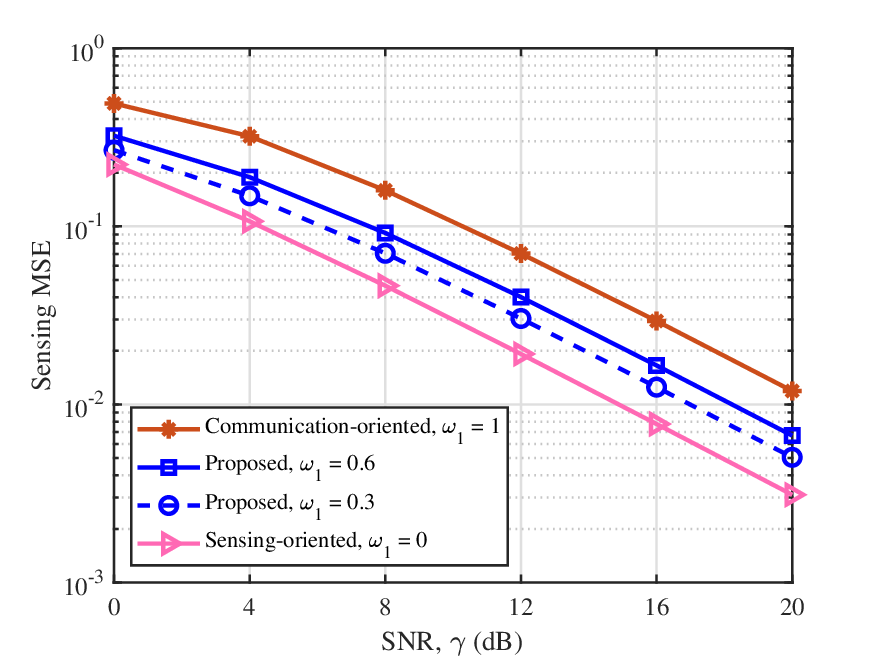}
%\caption{fig2}
\end{minipage}%
}%
\centering
\caption{ MSE performances of the statistical CSI-based design ($\gamma^\text{CE} = \gamma^\text{DT} = \gamma $).}\label{fig:sta_MSE}
\end{figure}

First, we verify the performance of our proposed sequential design with instantaneous CSI feedback.
Fig. \ref{fig:ins_PCE} evaluates the communication and sensing MSEs achieved by different schemes versus $ \gamma^\text{DT} $ with instantaneous CSI feedback, where $\gamma^\text{CE} = 1$~dB and $\omega_1 = 0.8$.
Firstly, we find that an increase of $\gamma^\text{DT}$ results in a larger power in the transmission stage, thereby reducing the communication and sensing MSEs. In addition, the communication-oriented and the sensing-oriented schemes can achieve the best data transmission and TRM estimation performances, respectively.
Secondly, compared to the existing scheme, the proposed scheme can significantly improve the TRM estimation performance as shown in Fig.~\ref{fig:ins_PCE}(b), due to the ISAC training signal optimization.
Moreover, although the additional sensing purpose of the training signal may affect the accuracy of communication CSI estimation, the data transmission performance of our proposed scheme is very close to that of the existing scheme, as illustrated in Fig.~\ref{fig:ins_PCE}(a). This is due to the robust design implemented in our work, which effectively mitigates the impact of CSI estimation errors on the communication performance.

Then, we evaluate the statistical CSI-based joint optimization scheme. Fig.~\ref{fig:sta_MSE} shows the communication and sensing MSEs, achieved by solving $(\mathcal P_3)$, versus the training and transmission SNR $\gamma^\text{CE} = \gamma^\text{DT} = \gamma $.
It is observed from the figures that both the sensing and communication performances are improved with the increase of $\gamma$.
On the other hand, when we gradually decrease the weighting factor $\omega_1$ from 1 to 0, corresponding to the four curves in the figures, the system is accordingly converted from communication-dominated to sensing-dominated. Therefore, it is seen that the communication MSE becomes large and, in the contrary, the sensing performance gets better gradually.

\begin{table}[t]
\caption{Comparison of Training and CSI Feedback Overheads}
    \centering
	\begin{tabular}{|c|c|c|}
\hline \textbf{Scheme} & \textbf{Training overhead} & \textbf{CSI feedback overhead} \\
\hline Sequential design & $M$ & $M N^\text{com}$ complex numbers \\
\hline Joint design & $M$ & - \\
\hline
	\end{tabular}\label{table:CSIcomparison}
\end{table}

In what follows, we compare the performances of the proposed instantaneous CSI-based design scheme and the statistical CSI-based scheme.
First, the training and feedback overheads of the two schemes are compared in Table~\ref{table:CSIcomparison}.
We find that these two schemes have identical training overheads of $M$ due to the instantaneous CSI estimation at the receiver. However, the sequential design requires feeding back the instantaneous CSI of size $M \times N^\text{com}$ to the transmitter, thereby incurring an additional feedback overhead.

\begin{figure}[t]
\begin{center}
      \epsfxsize=7.0in\includegraphics[scale=0.52]{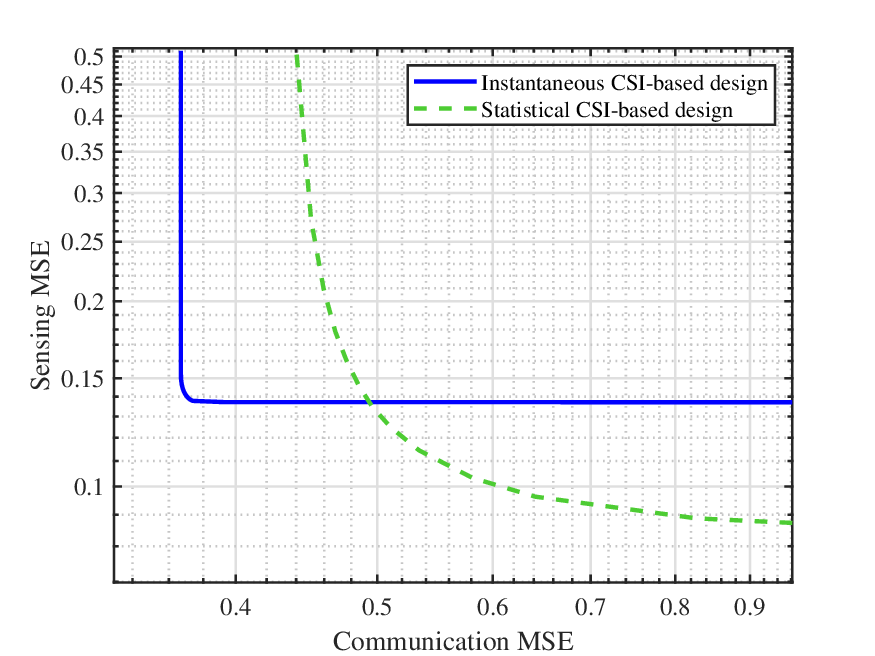}
      \caption{Communication-sensing MSE regions ($\gamma^\text{CE} = \gamma^\text{DT}  = 5$~dB).}\label{fig:region}
    \end{center}
\end{figure}

To proceed, we demonstrate the communication-sensing MSE regions. Specifically, the communication-sensing MSE region is utilized to characterize all the achievable MSE pairs for simultaneous communication and sensing under a given transmit power constraint, which is defined by
\begin{align}\label{def:MSEregion}
&\mathcal R_\text{com-rad} \triangleq \left\{ (  \widetilde{\text{MSE}}^\text{com},\widetilde{\text{MSE}}^\text{rad}):
\widetilde{\text{MSE}}^\text{com} \geq \frac{1}{D} \text{MSE}^\text{com}, \right. \nonumber \\
& \left. \widetilde{\text{MSE}}^\text{rad} \!\geq\! \frac{1}{M} \text{MSE}^\text{rad}, \text{Tr}\{ \mathbf X \mathbf X^H\} \!\leq\! P^\text{CE}, \text{Tr}\{ \mathbf W \mathbf W^H\} \!\leq\! P^\text{DT} \right\},
\end{align}
whose boundary can be characterized by consecutively adjusting the value of $\omega_1$ from 0 to 1 and solving the resulting problems, given in Sections~III and IV for instantaneous and statistical CSI-based schemes, respectively.
Fig. \ref{fig:region} illustrates the communication-sensing MSE region for the considered ISAC design. First, the tradeoff between communication and sensing is found from this figure, i.e., $\text{MSE}^\text{com}$ and $\text{MSE}^\text{rad}$ cannot decrease at the same time, since both functions share the limited power and spatial resources.
Moreover, it is seen that the instantaneous CSI-based scheme and the channel statistics-based scheme have the potential to achieve lower communication MSE and lower sensing MSE, respectively, which confirms to our expectations discussed at the end of Section~II.
More specifically, when considering a scenario with stringent radar requirements, we can apply the channel statistics-based design to achieve a better sensing MSE performance. While for a communication-dominant scenario, the instantaneous CSI-based design should be utilized to attain a better communication MSE performance.

\begin{figure}[t]
\centering
\subfigure[MSE region.]{
\begin{minipage}[t]{0.9\linewidth}
\centering
\includegraphics[scale=0.52]{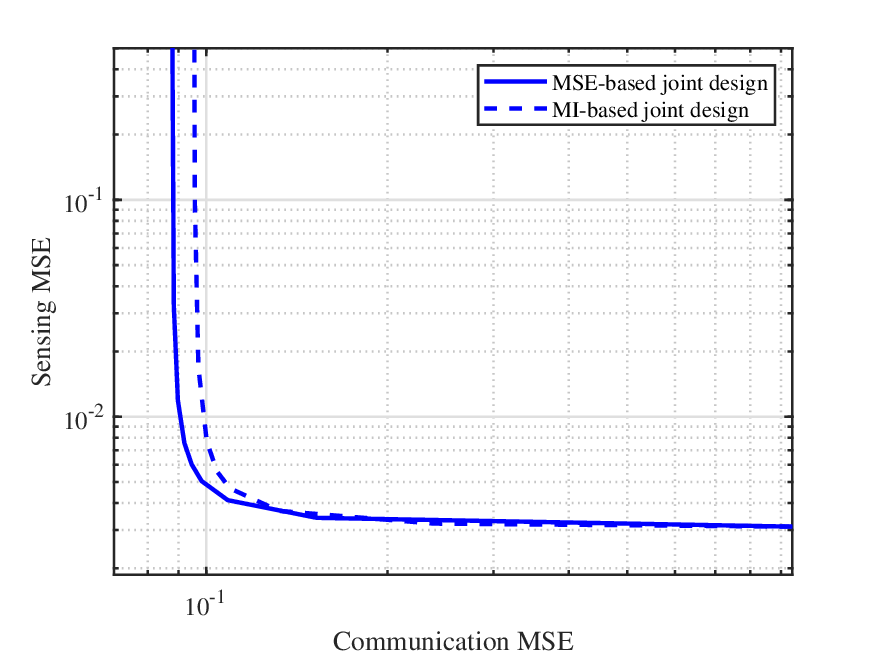}
%\caption{fig1}
\end{minipage}%
}%
\\
\subfigure[MI region.]{
\begin{minipage}[t]{0.9\linewidth}
\centering
\includegraphics[scale=0.52]{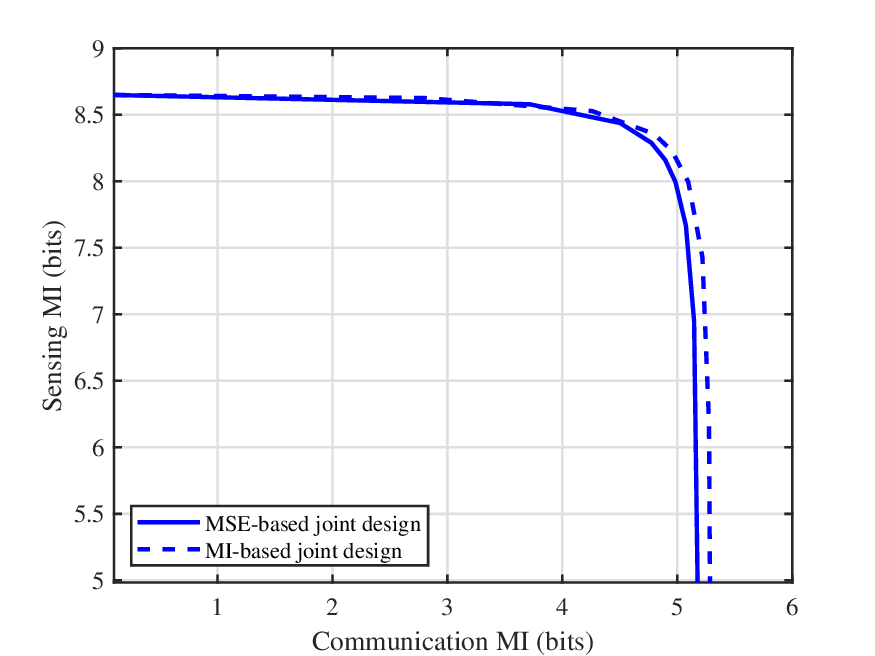}
%\caption{fig2}
\end{minipage}%
}%
\centering
\caption{Communication-sensing MSE and MI regions of the MSE-based joint design and the MI-based joint design ($\gamma^\text{CE} = \gamma^\text{DT} = 20$~dB).}\label{fig:region_MI}
\end{figure}

Finally, we compare the achievable performances of the MSE-based and the MI-based joint designs, by presenting the communication-sensing MSE and MI regions in Fig.~\ref{fig:region_MI}, where the definition of the MI region is similar to (\ref{def:MSEregion}) and is omitted.
We observe from Fig.~\ref{fig:region_MI}(a) that, compared to the MI-based design, the MSE-based design has the potential to achieve lower communication MSE, which is as expected since its design goal is MSE minimization.
At the same time, a larger communication MI can be achieved by the MI-based criterion as shown in Fig.~\ref{fig:region_MI}(b). On the other hand, we focus on the extreme scenario where $\omega_1 \rightarrow 0$ and the system tends to be sensing-oriented, corresponding to the rightmost end of the curves in Fig.~\ref{fig:region_MI}(a) and the leftmost end of the curves in Fig.~\ref{fig:region_MI}(b). It is seen that the two criteria can achieve nearly identical sensing performances in terms of either MSE or MI. This is because, for the target TRM estimation, the MSE minimization criterion and the MI maximization criterion result in the same waveform solution, as analyzed and substantiated in reference \cite{Y.Yang2007}.

\section{Conclusion}
In this paper, we investigated the joint training and transmission design in MIMO ISAC systems, aiming to minimize the weighted sum MSEs of both data communication and sensing TRM estimation. We first considered a sequential design scheme to optimize the training and the transmission signals separately, which requires the instantaneous estimated CSI feedback.
We also considered a channel statistics-based scheme to jointly optimize the training and the transmission signals through solving one single problem, which can reduce the CSI feedback overhead.
In addition, structured and low-cost solutions under special cases are provided.
The superiorities of the proposed schemes, especially the radar performance enhancement, can be clearly demonstrated through the simulations.
In particular, it is confirmed that the proposed instantaneous CSI-based scheme and the channel statistics-based scheme have the potential to achieve lower communication MSE and lower sensing MSE, respectively. These outcomes can be correspondingly leveraged in scenarios where communication or sensing predominates.

\begin{appendices}
\section{Proof of Proposition~\ref{prop:MM_P2}}\label{proof:MM_P2}
We derive two upper bounds for the communication and sensing MSEs in the objective function of $(\mathcal P_2)$, respectively, which will be used by the proposed MM algorithm.
To begin with, for the first term $\text{MSE}^\text{com}$, we employ the matrix inversion lemma and obtain
\begin{align}\label{mse_com_temp}
&[ \mathbf I_{D} +  \mathbf W^H \mathbf{\hat H}^H ((\text{Tr}\{ \mathbf W \mathbf W^H \mathbf R_{\mathbf \Delta} \} + \sigma^2)\mathbf I_{N^\text{com}} )^{-1} \mathbf{\hat H} \mathbf W]^{-1} \nonumber \\
=&\   -  \mathbf W^H \mathbf{\hat H}^H [  \mathbf{\hat H} \mathbf W \mathbf W^H \mathbf{\hat H}^H+ (\text{Tr}\{ \mathbf W \mathbf R_{\mathbf \Delta} \} + \sigma^2)\mathbf I_{N^\text{com}} ]^{-1} \nonumber \\
&\ \quad\ \times  \mathbf{\hat H} \mathbf W + \mathbf I_{D}.
\end{align}
Based on the fact that the function $f(\mathbf A, \mathbf B) = \text{Tr}\{ \mathbf A \mathbf B^{-1} \mathbf A^H \}$ is jointly convex with respect to $\{\mathbf A, \mathbf B\}$ when $\mathbf B \succ \mathbf 0$ \cite{cvxbook}, we define $\mathbf Q \triangleq \mathbf{\hat H} \mathbf W \mathbf W^H \mathbf{\hat H}^H + (\text{Tr}\{ \mathbf W \mathbf W^H \mathbf R_{\mathbf \Delta} \} + \sigma^2)\mathbf I_{N^\text{com}} \succ \mathbf 0$ and convert (\ref{mse_com_temp}) to $\text{MSE}^\text{com}(\mathbf W,\mathbf Q) = \text{Tr}\{  \mathbf I_{D} -  \mathbf W^H \mathbf{\hat H}^H \mathbf Q^{-1} \mathbf{\hat H} \mathbf W \}$, which is jointly concave with respect to $\{\mathbf W,\mathbf Q\}$.
Then, we apply the first-order Taylor expansion to $\text{MSE}^\text{com}(\mathbf W,\mathbf Q)$ and obtain an upper bound as follows:
\begin{align}\label{P3_up1_Taylor}
&\  \text{Tr}\left\{  \mathbf I_{D} -  \mathbf W^H \mathbf{\hat H}^H \mathbf Q^{-1} \mathbf{\hat H} \mathbf W \right\} \nonumber\\
\leq &  -\!\! \text{Tr}\left\{ \mathbf W^H_0 \mathbf{\hat H}^H \mathbf Q^{-1}_0 \mathbf{\hat H} \mathbf W_0 \!+\! 2 \mathcal R\{ \mathbf W^H_0 \mathbf{\hat H}^H \mathbf Q^{-1}_0 \mathbf{\hat H} ( \mathbf W \!-\! \mathbf W_0)\}  \right.  \nonumber\\
& \left. \quad\quad\quad - \mathbf Q_0^{-1} \mathbf{\hat H} \mathbf W_0 \mathbf W^H_0 \mathbf{\hat H}^H \mathbf Q_0^{-1}(\mathbf Q - \mathbf Q_0)
\right\} + D \nonumber\\
= &\ \text{Tr} \left\{  2 \mathcal R\{ \mathbf W^H_0 \mathbf{\hat H}^H \mathbf Q^{-1}_0 \mathbf{\hat H} \mathbf W \} \right\} \nonumber \\
&\ - \text{Tr}\left\{  \mathbf Q_0^{-1} \mathbf{\hat H} \mathbf W_0 \mathbf W^H_0 \mathbf{\hat H}^H \mathbf Q_0^{-1}\mathbf Q\right\} +  D\nonumber \\
\triangleq &\ \text{MSE}(\mathbf W,\mathbf Q;\mathbf W_0, \mathbf Q_0),
\end{align}
where $\mathbf Q_0  = \mathbf{\hat H} \mathbf W_0 \mathbf W_0^H \mathbf{\hat H}^H + (\text{Tr}\{ \mathbf W_0 \mathbf W_0^H \mathbf R_{\mathbf \Delta} \} + \sigma^2)\mathbf I_{N^\text{com}}$ and $\mathbf W_0$ is an arbitrary feasible point. Then, we substitute the definition of $\mathbf Q$ into $\text{MSE}(\mathbf W,\mathbf Q; \mathbf W_0, \mathbf Q_0)$, yielding
\begin{align}\label{P3_up1}
 &\ \text{MSE}(\mathbf W,\mathbf Q; \mathbf W_0, \mathbf Q_0)  \nonumber\\
=&\ D
- \text{Tr}\left\{  2 \mathcal R\{ \mathbf W^H_0 \mathbf{\hat H}^H \mathbf Q^{-1}_0 \mathbf{\hat H} \mathbf W \}\right\} \nonumber\\
&+ \text{Tr}\left\{ \mathbf Q_0^{-1} \mathbf{\hat H} \mathbf W_0 \mathbf W^H_0 \mathbf{\hat H}^H \mathbf Q_0^{-1}\mathbf{\hat H} \mathbf W \mathbf W^H \mathbf{\hat H}^H \right\} \nonumber \\
& + \text{Tr}\{ \mathbf W \mathbf W^H \mathbf R_{\mathbf \Delta} \}
\text{Tr}\left\{ \mathbf Q_0^{-1} \mathbf{\hat H} \mathbf W_0 \mathbf W^H_0 \mathbf{\hat H}^H \mathbf Q_0^{-1}
\right\} \nonumber\\
 &+  \sigma^2
\text{Tr}\left\{ \mathbf Q_0^{-1} \mathbf{\hat H} \mathbf W_0 \mathbf W^H_0 \mathbf{\hat H}^H \mathbf Q_0^{-1}
\right\} \nonumber \\
=&\ \text{Tr}\left\{ \mathbf W^H \mathbf \Psi \mathbf W \right\} \!-\! \text{Tr}\left\{  2 \mathcal R\{ \mathbf W^H_0 \mathbf{\hat H}^H \mathbf Q^{-1}_0 \mathbf{\hat H} \mathbf W \}\right\} \!+\! c_1,
\end{align}
where $c_1$ is the constant term and $\mathbf \Psi$ is given by
\begin{align}\label{def:Psi}
 \mathbf \Psi \triangleq&\  \mathbf{\hat H}^H \mathbf Q_0^{-1} \mathbf{\hat H} \mathbf W_0 \mathbf W^H_0 \mathbf{\hat H}^H \mathbf Q_0^{-1}\mathbf{\hat H} \nonumber\\
& + \text{Tr}\left\{ \mathbf Q_0^{-1} \mathbf{\hat H} \mathbf W_0 \mathbf W^H_0 \mathbf{\hat H}^H \mathbf Q_0^{-1}\right\}   \mathbf R_{\mathbf \Delta} \succeq \mathbf 0.
\end{align}

In a similar way, we manipulate $\text{MSE}^\text{rad}$ in the objective function of $(\mathcal P_2)$ in the following. First,
by defining $\mathbf R_{\mathbf G'} \triangleq \mathbf R_{\mathbf G}^{-1} + \frac{1}{\sigma^2} \mathbf X^\star(\mathbf X^\star)^H$ for notational ease and employing the matrix inversion lemma, we have
\begin{align}
& \text{MSE}^\text{rad}\! =\! \text{Tr}\left\{ \mathbf R_{\mathbf G'}^{-1} \right\}  \!-\! \text{Tr}\left\{  \mathbf R_{\mathbf G'}^{-1} \mathbf W \left(  \sigma^2 \mathbf I_D/L^\text{DT} \!\!+\!\! \mathbf W^H  \mathbf R_{\mathbf G'}^{-1} \mathbf W \right)^{-1} \right.\nonumber\\
& \left. \quad\quad\quad\quad\quad\quad\quad\quad\quad\quad\quad \times \mathbf W^H  \mathbf R_{\mathbf G'}^{-1} \right\}.
\end{align}
To proceed, defining $ \mathbf \Xi \triangleq \frac{\sigma^2}{L^\text{DT}} \mathbf I_D + \mathbf W^H  \mathbf R_{\mathbf G'}^{-1} \mathbf W \succ \mathbf 0$, we have
\begin{align}\label{P3_up2}
&\ \text{MSE}^\text{rad} \nonumber\\
\leq &\ \text{Tr}\left\{ \mathbf \Xi_0^{-1} \mathbf W^H_0  \mathbf R_{\mathbf G'}^{-2} \mathbf W_0 \mathbf \Xi_0^{-1} \mathbf W^H  \mathbf R_{\mathbf G'}^{-1} \mathbf W \right\} \nonumber\\
&\ -  \text{Tr}\left\{  2 \mathcal R\{  \mathbf \Xi_0^{-1}\mathbf W_0^H \mathbf R_{\mathbf G'}^{-2} \mathbf W   \} \right\}+ c_2 \nonumber\\
\leq &\  -  2\text{Tr}\left\{\mathcal R\{  \mathbf \Xi_0^{-1}\mathbf W_0^H \mathbf R_{\mathbf G'}^{-2} \mathbf W + \lambda \mathbf W_0^H \mathbf W \right. \nonumber \\
&\ \left.- \mathbf \Xi_0^{-1} \mathbf W^H_0  \mathbf R_{\mathbf G'}^{-2} \mathbf W_0 \mathbf \Xi_0^{-1} \mathbf W_0^H  \mathbf R_{\mathbf G'}^{-1} \mathbf W  \} \right\} \nonumber\\
&\ + \lambda \text{Tr}\{\mathbf W^H \mathbf W \}+ c_3.
\end{align}
In (\ref{P3_up2}), the first inequality follows from the first-order Taylor expansion on $\text{Tr}\left\{  \mathbf R_{\mathbf G'}^{-1} \mathbf W \mathbf \Xi^{-1} \mathbf W^H  \mathbf R_{\mathbf G'}^{-1} \right\}$, where $\mathbf \Xi_0 = \frac{\sigma^2}{L^\text{DT}} \mathbf I_D + \mathbf W_0^H  \mathbf R_{\mathbf G'}^{-1} \mathbf W_0$, the second inequality holds due to \cite[Eq. (26)]{Y.SunTSP2017MM}, where $\lambda$ can be calculated as
\begin{align}\label{def:lambda}
\lambda = \lambda_\text{max} (\mathbf \Xi_0^{-1} \mathbf W^H_0  \mathbf R_{\mathbf G'}^{-2} \mathbf W_0 \mathbf \Xi_0^{-1} ) \lambda_\text{max} ( \mathbf R_{\mathbf G'}^{-1}),
\end{align}
and $c_2$ and $c_3$ stand for the constant terms.
By combining the upper bounds obtained in (\ref{P3_up1}) and (\ref{P3_up2}) and removing the constant terms, we obtain the surrogate problem in (\ref{prob:p2MM}), where
\begin{align}\label{def:Pi}
\mathbf \Pi^H =&\ \frac{\omega_1}{D} \mathbf W^H_0 \mathbf{\hat H}^H \mathbf Q^{-1}_0 \mathbf{\hat H} +
\frac{\omega_2}{M} (\mathbf \Xi_0^{-1}\mathbf W_0^H \mathbf R_{\mathbf G'}^{-2}  +  \lambda \mathbf W_0^H \nonumber\\
&\ - \mathbf \Xi_0^{-1} \mathbf W^H_0  \mathbf R_{\mathbf G'}^{-2} \mathbf W_0 \mathbf \Xi_0^{-1} \mathbf W_0^H  \mathbf R_{\mathbf G'}^{-1} ).
\end{align}

\section{Proof of Theorem~\ref{theorem:MMconvergence}}\label{proof:MMconvergence}
Let us denote the objective function and the constraint set of the original problem $(\mathcal P_1)$ by $f(\mathbf W)$ and $\mathcal S$, respectively.
The conditions required by the MM framework for constructing the surrogate function, denoted by $f(\mathbf W;\mathbf W_0)$, are listed as follows \cite[Section III]{MMAcce1}:
\begin{enumerate}
  \item $f\left(\mathbf W\right) \leq f\left(\mathbf W;\mathbf W_0\right), \ \forall \mathbf W \in \mathcal S$;
  \item $f\left( \mathbf W_0 \right) = f\left(\mathbf W_0; \mathbf W_0 \right)$;
  \item $\nabla f\left( \mathbf W_0 \right) = \nabla f\left( \mathbf W_0; \mathbf W_0 \right)$;
  \item $f\left(\mathbf W; \mathbf W_0\right)$ is continuous in both $\mathbf W$ and $\mathbf W_0$.
\end{enumerate}
When all the four conditions hold, the MM-based iterative algorithm can be guaranteed to converge to a stationary point of the original problem. In the sequel, we prove that the utilized surrogate function, i.e., the objective function of problem (\ref{prob:p2MM}), and the objective function of problem $(\mathcal P_2)$ satisfy these conditions.

Firstly, it is clearly seen that the objective function of problem (\ref{prob:p2MM}) is a continuous function and thus condition 4) holds. Secondly, as shown in (\ref{P3_up1_Taylor}), the employed upper bound for the communication MSE is derived based on the first-order Taylor expansion, for which 1), 2), and 3) can be readily verified to hold. On the other hand, we utilize the first-order and the second-order Taylor expansions in (\ref{P3_up2}) to upper bound the radar sensing MSE, where conditions 1), 2), and 3) also hold. Hence, all the four conditions hold for the surrogate function in (\ref{prob:p2MM}) and the convergence proof is completed.

\section{Proof of Proposition~\ref{prop:prob_equ}}\label{proof:prob_equ}
By substituting $\mathbf R_{\mathbf X} = \mathbf X \mathbf X^H$, the subproblem of the training signal $\mathbf X$ becomes
\begin{align}\label{prob:R2_original}
\mathop \text{minimize}\limits_{\mathbf R_{\mathbf X}} \quad &
\frac{\omega_1}{D} \text{Tr}\left\{ \left( \mathbf I_{D} +  \frac{N^\text{com} \mathbf W^H \mathbf R_\mathbf{\hat H} \mathbf W} {\text{Tr}\{ \mathbf W \mathbf W^H \mathbf R_{\mathbf \Delta} \} + \sigma^2} \right)^{-1}\right\} \nonumber \\
& + \frac{\omega_2 }{M} f(\mathbf R_{\mathbf X}) \nonumber \\
\text{subject to} \quad & \mathbf R_{\mathbf \Delta} = \left(\mathbf R_{\mathbf H}^{-1} + \frac{1}{\sigma^2} \mathbf R_{\mathbf X} \right)^{-1}, \
 \mathbf R_{\mathbf {\hat H}} = \mathbf R_{\mathbf H} - \mathbf R_{\mathbf \Delta}, \nonumber\\
 & \text{Tr}\left\{ \mathbf R_{\mathbf X}\right\} \leq P^\text{CE},\ \mathbf R_{\mathbf X} \succeq \mathbf 0,
\end{align}
where $f(\mathbf R_{\mathbf X}) \triangleq \text{Tr}\left\{ \left(\mathbf R_{\mathbf G}^{-1} + \frac{1}{\sigma^2} \mathbf R_{\mathbf X} + \frac{L^\text{DT}}{\sigma^2} \mathbf W \mathbf W^H \right)^{-1} \right\}$ denotes the sensing MSE and it is convex with respect to $\mathbf R_{\mathbf X}$. The remaining issue of solving (\ref{prob:R2_original}) is to deal with the nonconvex communication MSE.

To begin with, we introduce a positive semidefinite matrix $\mathbf \Gamma$ as an auxiliary variable and transform problem (\ref{prob:R2_original}) into the following equivalent form:
\begin{align}\label{prob:R2_Gamma}
\mathop \text{minimize}\limits_{\mathbf R_{\mathbf X},\mathbf \Gamma } \quad &
\frac{\omega_1}{D} \text{Tr}\left\{ \left( \mathbf I_{D} +  \frac{N^\text{com} \mathbf W^H (\mathbf R_{\mathbf H} - \mathbf \Gamma) \mathbf W} {\text{Tr}\{ \mathbf W \mathbf W^H \mathbf \Gamma \} + \sigma^2} \right)^{-1}\right\}
\nonumber \\
& + \frac{\omega_2 }{M} f(\mathbf R_{\mathbf X}) \nonumber \\
\text{subject to} \quad & \mathbf \Gamma \succeq \left(\mathbf R_{\mathbf H}^{-1} + \frac{1}{\sigma^2} \mathbf R_{\mathbf X} \right)^{-1}, \
 \text{Tr}\left\{ \mathbf R_{\mathbf X}\right\} \leq P^\text{CE},\nonumber \\
& \mathbf R_{\mathbf X} \succeq \mathbf 0,\
  \mathbf \Gamma \succeq \mathbf 0.
\end{align}
The equivalence is established based on the fact that $\mathbf \Gamma = \left(\mathbf R_{\mathbf H}^{-1} + \frac{1}{\sigma^2} \mathbf R_{\mathbf X} \right)^{-1}$ must hold at the optimality, which is proved by contradiction.
Specifically, assume that $\mathbf \Gamma'$ is an optimal solution of (\ref{prob:R2_Gamma}), where $\mathbf \Gamma' \succeq \left(\mathbf R_{\mathbf H}^{-1} + \frac{1}{\sigma^2} \mathbf R_{\mathbf X} \right)^{-1}$ and $\mathbf \Gamma' \neq \left(\mathbf R_{\mathbf H}^{-1} + \frac{1}{\sigma^2} \mathbf R_{\mathbf X} \right)^{-1}$. In such case, we can always find a matrix $\hat {\mathbf \Gamma}$ satisfying $\mathbf \Gamma' \succeq \hat {\mathbf \Gamma} \succeq \left(\mathbf R_{\mathbf H}^{-1} + \frac{1}{\sigma^2} \mathbf R_{\mathbf X} \right)^{-1}$ and $\hat {\mathbf \Gamma} \neq \mathbf \Gamma'$. It follows that $\text{Tr}\{ \mathbf W \mathbf W^H \hat{\mathbf \Gamma} \} < \text{Tr}\{ \mathbf W \mathbf W^H \mathbf \Gamma' \}$ and $\mathbf W^H (\mathbf R_{\mathbf H} - \hat{\mathbf \Gamma}) \mathbf W \succeq \mathbf W^H (\mathbf R_{\mathbf H} - \mathbf \Gamma') \mathbf W $ \cite{book:MatrixAnalysis}. Hence, replacing $\mathbf \Gamma'$ with $\hat {\mathbf \Gamma}$ results in a smaller objective, which contradicts with the assumption that $\mathbf \Gamma'$ is optimal. Therefore, $\mathbf \Gamma = \left(\mathbf R_{\mathbf H}^{-1} + \frac{1}{\sigma^2} \mathbf R_{\mathbf X} \right)^{-1}$ must hold at the optimality and, hence, problems (\ref{prob:R2_original}) and (\ref{prob:R2_Gamma}) are equivalent.

To proceed, we introduce a real-valued auxiliary variable $t$ to problem (\ref{prob:R2_Gamma}) and obtain the following equivalent problem:
\begin{align}\label{prob:R2_t}
\mathop \text{minimize}\limits_{\mathbf R_{\mathbf X},\mathbf \Gamma,t \geq 0  } \quad &
\frac{\omega_1}{D} \text{Tr}\left\{ \left( \mathbf I_{D} +  \frac{N^\text{com} \mathbf W^H (\mathbf R_{\mathbf H} - \mathbf \Gamma) \mathbf W} {t } \right)^{-1}\right\}
\nonumber \\
& + \frac{\omega_2 }{M} f(\mathbf R_{\mathbf X}) \nonumber \\
\text{subject to} \quad &  \text{constraints of problem (\ref{prob:R2_Gamma})}, \nonumber \\
& t \geq \text{Tr}\{ \mathbf W \mathbf W^H \mathbf \Gamma \} + \sigma^2.
\end{align}
The equivalence holds since the constraint $t \geq \text{Tr}\{ \mathbf W \mathbf W^H \mathbf \Gamma \} + \sigma^2$ must keep active at the optimality, which can be proved by contradiction similarly.
Subsequently, based on the equality
$
\text{Tr}\left\{ \left( \mathbf I_{D} +  \frac{N^\text{com} \mathbf W^H (\mathbf R_{\mathbf H} - \mathbf \Gamma) \mathbf W} {t } \right)^{-1}\right\}
=
t
\text{Tr}\left\{ \left( t \mathbf I_{D} + N^\text{com} \mathbf W^H (\mathbf R_{\mathbf H} - \mathbf \Gamma) \mathbf W \right)^{-1}\right\},
$
we introduce another real-valued auxiliary variable $u$ and rewrite (\ref{prob:R2_t}) by the following equivalent form:
\begin{align}
\mathop \text{minimize}\limits_{\mathbf R_{\mathbf X},\mathbf \Gamma,t \geq 0, u\geq 0  } \quad &
\frac{\omega_1}{D} tu
+ \frac{\omega_2 }{M} f(\mathbf R_{\mathbf X}) \nonumber \\
\text{subject to} \quad &  \text{constraints of problem (\ref{prob:R2_t})}, \nonumber \\
& u \geq \text{Tr}\left\{ \left( t \mathbf I_{D} + N^\text{com} \mathbf W^H (\mathbf R_{\mathbf H} - \mathbf \Gamma) \mathbf W \right)^{-1}\right\}, \nonumber
\end{align}
which corresponds to problem (\ref{prob:JD_Xtu}).

\section{Proof of Proposition~\ref{prop:X_P1}}\label{proof:X_P1}
By substituting the EVDs $\mathbf R_{\mathbf H} = \mathbf U \mathbf \Lambda_{\mathbf H} \mathbf U^H$ and $\mathbf R_{\mathbf G} = \mathbf U \mathbf \Lambda_{\mathbf G} \mathbf U^H$ into the objective function of $(\mathcal P_1)$, we have
\begin{align}\label{MSE:U}
&\ \text{Tr}\{(\mathbf R_{\mathbf H}^{-1} \!+\! \mathbf X \mathbf X^H/\sigma^2 )^{-1} \}
= \text{Tr}\{(\mathbf \Lambda_{\mathbf H}^{-1} \!+\! \mathbf U^H \mathbf X \mathbf X^H \mathbf U/\sigma^2 )^{-1} \}, \nonumber\\
&\ \text{Tr}\{(\mathbf R_{\mathbf G}^{-1} \!+\! \mathbf X \mathbf X^H/\sigma^2)^{-1} \}
= \text{Tr}\{(\mathbf \Lambda_{\mathbf G}^{-1} \!+\! \mathbf U^H\mathbf X \mathbf X^H \mathbf U/\sigma^2 )^{-1}\}.
\end{align}
To proceed, we introduce the following lemma.

\begin{lemma}[\cite{M.BigueshTSP2006Training-based}]
For an arbitrary $N \times N$ positive definite matrix $\mathbf A$, it holds that $ \text{Tr}\{ \mathbf A^{-1}\} \geq \sum_{n=1}^N [\mathbf A]_{n,n}^{-1}$, where $[\mathbf A]_{n,n}$ stands for the $n$-th diagonal element of $\mathbf A$. The equality holds if $\mathbf A$ is diagonal.
\end{lemma}\label{lemma:1}

Based on Lemma~1, we conclude that, in order to minimize the MSEs in (\ref{MSE:U}), $\mathbf U^H \mathbf X \mathbf X^H \mathbf U$ must be a diagonal matrix, i.e., $\mathbf X$ has the structure as $\mathbf X = \mathbf U \mathbf \Lambda_{\mathbf X}$.
Furthermore, by substituting $\mathbf X = \mathbf U \mathbf \Lambda_{\mathbf X}$ into $(\mathcal P_1)$, we obtain the following equivalent problem with respect to the diagonal matrix $\mathbf \Lambda_{\mathbf X}$:
\begin{align}\label{prob:p}
\!\!\mathop \text{minimize}\limits_{ \{ x_m \geq 0\}_{m=1}^{M}} \  &
\frac{\omega_1}{M} \!\! \sum_{m=1}^{M}\! \!(\lambda_{\mathbf H,m}^{-1} \!\!+\! x_m/\sigma^2 )^{-1}
 \!\!+\!\! \frac{\omega_2}{M} \!\! \sum_{m=1}^{M}\!\! (\lambda_{\mathbf G,m}^{-1} \!\!+\! x_m/\sigma^2 )^{-1} \nonumber \\
\text{subject to} \  & \sum_{m=1}^{M}  x_m \leq P^\text{CE}.
\end{align}
This is a convex problem with respect to $\{ x_m\}_{m=1}^M$, which can also been addressed employing the Lagrange multiplier method \cite{cvxbook}. To this end, denoting the lagrange multiplier associated with the power constraint by $\mu \geq 0$, we then give the partial Lagrange function of problem (\ref{prob:p}) by
\begin{align}
&\ \mathcal L( \{ x_m\}_{m=1}^{M}, \mu) \nonumber \\
=&\ \frac{\omega_1}{M} \sum_{m=1}^{M} \left(\lambda_{\mathbf H,m}^{-1} \!+\! x_m/\sigma^2\right)^{-1}
\!\!+\!\!
\frac{\omega_2}{M} \sum_{m=1}^{M}\left(\lambda_{\mathbf G,m}^{-1} \!+\! x_m/\sigma^2\right)^{-1}\nonumber \\
&\ + \mu \left(\sum_{m=1}^{M} x_m - P^\text{CE} \right).
\end{align}
By solving $\frac{\partial \mathcal L}{\partial x_m} = 0$ for each $m$, we obtain the condition in (\ref{xm_P1}).
Since $x_m \geq 0$, from (\ref{xm_P1}) we have $\mu^\star \neq 0$ and obtain the upper bound of $\mu^\star$ in (\ref{def:hat_mu}).
In addition, the complementary slackness condition of problem (\ref{prob:p}) is given by
$\mu^\star (\sum_{m=1}^{M} x_m^\star  - P^\text{CE}) = 0.$
Based on $\mu^\star > 0$, we have $\sum_{m=1}^{M}  x_m^\star = P^\text{CE}$. The proof is completed.

\section{Proof of Theorem~\ref{theorem:structureofP3}}\label{proof:structureofP3}
Utilizing Lemma~1 shown in Appendix~\ref{proof:X_P1}, it is verified that (\ref{struture:p3}) is a sufficient and necessary condition of minimizing the second term of the objective function of $(\mathcal P_3)$, i.e., $\text{MSE}^\text{rad}$. Therefore, the remaining issue is to prove that (\ref{struture:p3}) is also optimal for minimizing the first term, i.e., $\overline{\text{MSE}}^\text{com} = \text{Tr}\left\{ \left( \mathbf I_{D} +  \frac{ N^\text{com} \mathbf W^H \mathbf R_\mathbf{\hat H} \mathbf W} { \text{Tr}\{ \mathbf W \mathbf W^H \mathbf R_{\mathbf \Delta}\} + \sigma^2 } \right)^{-1}\right\}$, which is given as follows.

Define $\kappa \triangleq \text{Tr}\{ \mathbf W \mathbf W^H \mathbf R_{\mathbf \Delta}\} + \sigma^2 > 0$ and $\overline{\text{MSE}}^\text{com}$ becomes $\text{Tr}\{ ( \mathbf I_{D} + \frac{1}{\kappa} \mathbf W^H \mathbf R_\mathbf{\hat H} \mathbf W )^{-1}\}$.
Denote the EVD of $\mathbf R_\mathbf{\hat H}$ and $\mathbf R_\mathbf{\Delta}$ by $\mathbf R_\mathbf{\hat H} = \mathbf U_\mathbf{\hat H} \mathbf \Lambda_\mathbf{\hat H} \mathbf U_\mathbf{\hat H}^H$ and $\mathbf R_\mathbf{\Delta} = \mathbf U_\mathbf{\Delta} \mathbf \Lambda_\mathbf{\Delta} \mathbf U_\mathbf{\Delta}^H$, respectively.
Then, we obtain from Lemma~1 that the optimal $\mathbf W$ that minimizes $\overline{\text{MSE}}^\text{com}$ with a fixed $\kappa$ obeys $\mathbf W = \mathbf U_\mathbf{\hat H} \mathbf \Lambda_{\mathbf W}$.
In other words, the directions of $\mathbf W$ should be aligned with those of the correlation matrix of the estimated channel. Meanwhile, for the purpose of minimizing $ \kappa$, $\mathbf W$ should satisfy $\mathbf W = \mathbf U_\mathbf{\Delta} \mathbf \Lambda_{\mathbf W}$ \cite{book:mtxMajorization}.
Moreover, when $\mathbf X = \mathbf U \mathbf \Lambda_{\mathbf X}$, it can be readily shown that $\mathbf R_\mathbf{\hat H}$ and $\mathbf R_\mathbf{\Delta}$ share the same eigenvectors, i.e., $\mathbf U_\mathbf{\hat H}= \mathbf U_\mathbf{\Delta}= \mathbf U$.
Therefore, the $\mathbf W $ and $\mathbf X$ given in (\ref{struture:p3}) also minimize $\overline{\text{MSE}}^\text{com}$. The proof is completed.

\section{Proof of Equivalence between \\Problems (\ref{prob:JD_PA}) and (\ref{prob:JD_PA_auxi})}\label{proof:GP_equ}
We first introduce $ \xi_{d} \geq 0, \ d\in \mathcal D,$ and transform (\ref{prob:JD_PA}) into the following equivalent problem:
\begin{align}\label{prob:R2_PA1}
\mathop \text{minimize}\limits_{\{x_m \}_{m=1}^M, \atop \{w_d,\xi_{d} \}_{d=1}^D} \quad &
\frac{\omega_1}{D} \sum_{d=1}^{D} \left( 1+ \frac{N^\text{com}}{\sum_{i=1}^D \xi_{d} + \sigma^2}
\frac{\lambda_{\mathbf H,d}^2 w_d x_d}{ \lambda_{\mathbf H,d}x_d + \sigma^2 } \right)^{-1} \nonumber\\
&+ \frac{\omega_2}{M}
 \sum_{d=1}^{D} \left(\lambda_{\mathbf G,d}^{-1} + \frac{1}{\sigma^2 } x_d + \frac{L^\text{DT}}{\sigma^2 } w_d \right)^{-1} \nonumber\\
& + \frac{\omega_2}{M}
 \sum_{m=D+1}^{M} \left(\lambda_{\mathbf G,m}^{-1} + \frac{1}{\sigma^2 } x_m\right)^{-1} \nonumber \\
\text{subject to} \quad & \text{constraints of problem (\ref{prob:JD_PA}},\nonumber \\
& \xi_d \geq \frac{\lambda_{\mathbf H,d}\sigma^2w_d}{\lambda_{\mathbf H,d}x_d +\sigma^2 }, \ d \in \mathcal D.
\end{align}
The equivalence holds since the objective function is monotonically increasing with $\{ \xi_{d}\}$ and thus the added inequalities must keep active at the optimality.
To proceed, we introduce auxiliary variables $t \geq 0$ and $\{\kappa^\text{rad}_{m}\geq 0 \}_{m=1}^M$, and transform (\ref{prob:R2_PA1}) into the following equivalent problem:
\begin{align}\label{prob:R2_PA2}
\mathop \text{minimize}\limits_{\{x_m,\kappa^\text{rad}_{m}\}_{m=1}^M,t, \atop \{w_d ,\xi_{d}  \}_{d=1}^D} \quad &
\frac{\omega_1}{D} \sum_{d=1}^{D} \left( 1+ \frac{N^\text{com}}{t}
\frac{\lambda_{\mathbf H,d}^2 w_d x_d}{ \lambda_{\mathbf H,d}x_d + \sigma^2 } \right)^{-1} \nonumber\\
&+ \frac{\omega_2}{M}
 \sum_{d=1}^{D} \left(\kappa^\text{rad}_{d} \right)^{-1}
+ \frac{\omega_2}{M}
 \sum_{m=D+1}^{M} \left(\kappa^\text{rad}_{m}\right)^{-1} \nonumber \\
\text{subject to} \quad & \text{constraints of problem (\ref{prob:R2_PA1})},\nonumber \\
& t \geq \sum_{d=1}^D \xi_{d} + \sigma^2, \nonumber \\
& \kappa_m^\text{rad} \leq \lambda_{\mathbf G,m}^{-1} + \frac{1}{\sigma^2 } x_m + \frac{L^\text{DT}}{\sigma^2 } w_m,  \  m \in \mathcal D \nonumber\\
& \kappa_m^\text{rad} \leq \lambda_{\mathbf G,m}^{-1} + \frac{1}{\sigma^2 } x_m,  \  m \in \mathcal D\backslash \mathcal M.
\end{align}
It can be proved by contradiction that the added constraints must be active at the optimality, and thus the equivalence between problems (\ref{prob:R2_PA1}) and (\ref{prob:R2_PA2}) holds.
Finally, we introduce $\{\kappa_d^\text{com} \geq 0 \}_{d=1}^D$ to similarly handle the first term in the objective function of problem (\ref{prob:R2_PA2}) and obtain problem (\ref{prob:JD_PA_auxi}).

\end{appendices}

\end{document}